%% file: y6kp_baosample_des.tex
\documentclass[
superscriptaddress,
twocolumn,
preprintnumbers,
nofootinbib,
 amsmath,amssymb,
 aps,prd,
floatfix,
]{revtex4-2}

\usepackage{graphicx}
\usepackage{dcolumn}
\usepackage{bm}
\usepackage[colorlinks,linkcolor=blue,citecolor=blue,urlcolor=blue ]{hyperref}
\usepackage{amsmath,amssymb,natbib,latexsym}
\usepackage[T1]{fontenc}
\usepackage[utf8]{inputenc}

\usepackage{hhline}
\usepackage{multirow}
\usepackage{xspace}

\usepackage{siunitx}

\newcommand{\code}[1]{\texttt{#1}}

\newcommand{\cosmolike}{\textsc{CosmoLike}\xspace}

\newcommand{\healpix}{\textsc{HEALPix}\xspace}
\newcommand{\zmean}{\textsc{DNF\_Z}\xspace}
\newcommand{\zmc}{\textsc{DNF\_ZN}\xspace}
\newcommand{\zspec}{\textsc{Z\_SPEC}\xspace}
\newcommand{\gold}{Y6 Gold\xspace}

\newcommand{\fitvd}{\code{fitvd}\xspace}
\newcommand{\sextractor}{\textsc{SExtractor}\xspace}
\newcommand{\flagsfootprint}{\textsc{FLAGS\_FOOTPRINT}\xspace}
\newcommand{\flagsgold}{\textsc{FLAGS\_GOLD}\xspace}

\newcommand{\extmash}{\textsc{EXT\_MASH}\xspace}
\newcommand{\nimages}[1][]{\textsc{N\_IMAGES}\_[{#1}]\xspace}
\newcommand{\bdft}{\textsc{BDF\_T}\xspace}
\newcommand{\bdfsn}{\textsc{BDF\_S2N}\xspace}
\newcommand{\nitermodel}{\textsc{NITER\_MODEL}\xspace}

\usepackage{array}
\newcolumntype{Z}{>{\setbox0=\hbox\bgroup}c<{\egroup}@{}}

\usepackage{booktabs}

\usepackage{xcolor}

\DeclareSIUnit\angstrom{\text {Å}}

\newcommand{\appendixcite}[1]{\hyperref[#1]{\textcolor{blue}{Appendix \ref*{#1}}}}

\begin{document}

\preprint{DES-2023-0797}
\preprint{FERMILAB-PUB-24-0072-PPD}

\title{Dark Energy Survey: Galaxy Sample for the Baryonic Acoustic Oscillation Measurement from the Final Dataset}

\input{authorlist}

\date{\today}

\begin{abstract}
    \input{abstract}
\end{abstract}

\maketitle


\section{\label{sec:intro}Introduction}

Baryon Acoustic Oscillations (BAO) is one of the most remarkable predictions of the formation of structures in the Universe \cite{Peebles1970, SZ1970, Bond1984, Bond1987}. Since its first detection in 2005 \cite{Eisenstein2005}, the measurement of the BAO scale has been one of the most important probes of dark energy and also one of the main scientific drivers in the design and construction of galaxy surveys.

The BAO signal has already been detected many times in spectroscopic \cite{Cole:2005, Percival:2007, Gaztanaga:2009, Percival:2010, Beutler:2011, Blake:2011, Ross:2015, Alam2017, bautista2021completed, gil2020completed, de2021completed, hou2021completed, neveux2020completed, alam2021completed} and photometric \cite{Padmanabhan:2007, Estrada:2009, Hutsi:2010, Crocce:2011, carnero2012, y1bao, y3bao} datasets for galaxies, but also in the distribution of QSOs \cite{Ata2018} and Lyman-$\alpha$ absorbers \cite{Bautista2017, des2020completed}, in a wide variety of redshifts, from $z=0.2$ to $z<3$. The estimation of the evolution of the BAO scale with time is a direct measurement of the expansion history of the Universe and, therefore, an excellent cosmology observable. All these measurements are compatible with the $\Lambda$CDM cosmological model.

In this context, the Dark Energy Survey (DES) \cite{Flaugher:2005, DES:2016ktf} aims to measure the BAO scale in the distribution of galaxies as one of its main objectives. In the DES Year 1 (Y1) analysis \cite{y1bao}, we measured the BAO scale at an effective redshift of 0.81 with a sample that covered 1,336 deg${}^2$. Because of this limited area, the detection had a low significance. On the other hand, in the DES Year 3 (Y3) analysis \cite{y3bao} we measured the BAO scale at an effective redshift of 0.835. The Y3 sample had a total of 7,031,993 galaxies and covered 4,108.47 deg${}^2$. Unlike in the case of the Y1, the significance of the detection in the Y3 was of about $3\sigma$ in the determination of the BAO feature. The BAO distance measurement obtained was $D_M(z_{\rm eff}=0.835)/r_{\rm d}=18.92\pm0.51$ (where $D_M(z)$ is the comoving angular diameter distance and $r_d$ is the sound horizon scale), making it the most precise BAO distance measurement from imaging data alone ever (2.7\% precision) and competitive with the latest transverse ones from spectroscopic samples at $z>0.75$. This result was consistent with Planck's prediction at the level of $2.3\sigma$. In the DES Year 6 (Y6) analysis, i.e., the dataset analyzed here, we expect to measure the BAO feature at an effective redshift of 0.867 with 25\% more precision compared to the Y3. Furthermore, this measurement will be combined with the other DES cosmological observables to estimate the most precise measurements on dark energy by the combination of BAO with 3$\times$2pt (galaxy clustering + weak lensing) and type Ia supernovae, similarly to what we did in the Y3 analysis \cite{y3bao}.

Detecting the BAO signal in photometric surveys poses a significant challenge due to the inherent smearing caused by the imprecise redshift determination. To mitigate this issue, it is crucial to identify a galaxy population exhibiting a distinctive spectral feature that can be captured using broadband filters. Generally, the preferred approach involves selecting old, well-evolved galaxies with a prominent 4,000 \SI{}{\angstrom} break \cite{Eisenstein:2001, Vakili:2019, y1baosample, Zhou:2020}. This characteristic imparts a reddish appearance to the galaxies and often serves as the primary criterion for target selection in galaxy surveys.

In \cite{y1baosample}, we developed a color selection to choose galaxies in the DES Y1 analysis, calibrated through a set of synthetic SED distributions, optimized for redshifts $z>0.5$. This same color selection was the one used for the DES Y3 analysis since we found it to be appropriate for the Y3 as well. 
In this new release, we re-optimize the Y1/Y3 sample selection. Also, since for the Y6 we have better data quality (deeper and more homogeneous) than for the Y3, i.e., less noisy magnitude estimations (because of the longer exposure time), we can afford to go deeper in magnitude (and also in redshift), which allows us to go up to $z_{\rm ph}=1.2$ (compared to the $z_{\rm ph}=1.0$ limit of the Y1 analysis, or the $z_{\rm ph}=1.1$ limit of the Y3).

The structure of the paper is as follows: in \autoref{sec:y6gold} we present the parent DES Y6 data and the Directional Neighborhood Fitting algorithm \cite{de2016dnf}, or DNF, which is the fiducial photo-$z$ code used within DES; in \autoref{sec:baosample} we describe the optimization of the DES Y6 BAO selection, together with the extra quality cuts we apply and its footprint; in \autoref{sec:photoz} we present the validation of the redshift distributions of our optimal Y6 sample, for which we perform a direct calibration with VIPERS and also compare with the results from clustering redshift (WZ) using SDSS galaxies (the fiducial redshift distributions used in our analysis are a combination of DNF, VIPERS and WZ);
in \autoref{sec:systematics} we describe the methodology used to mitigate the effect of observational systematics; in \autoref{sec:unblinded_clustering} we show the unblinded clustering measurements of our BAO sample;
and in \autoref{sec:conclusions} we show our conclusions to this analysis. The Y6 BAO sample will be eventually released at \url{https://des.ncsa.illinois.edu/releases}, together with all the other DES Y6 products.

In its companion paper \cite{Y6_BAO_measurement}, we measure the BAO scale as a function of redshift, using the sample optimized here. We run the analysis in configuration and Fourier spaces (using the $w(\theta)$ and $C_\ell$ statistics, respectively) and also using the projected correlation function (PCF) estimator (using the $\xi_p(s_\perp)$ statistics). Our fiducial measurement is the combination of the three estimators.


\section{DES Y6 Data}

The operations of the Dark Energy Survey ended in 2019, after six years of data-taking. DES used the Blanco 4m telescope at Cerro Tololo Inter-American Observatory (CTIO) in Chile and observed $\sim$5,000 deg${}^2$ of the southern sky in five broadband filters (bands), $grizY$, ranging from $\sim$400 nm to $\sim$1,060 nm \cite{Li:2016,Y3FGCM}, using the DECam \cite{Flaugher:2015} camera. Its images were processed with the DES Data Management system hosted at NCSA, coadded on colocated points in the sky for each band, from which catalogs of objects are produced using the combined detection in $riz$ bands \cite{desdm}. These final catalogs have been released as the public Data Release 2 of the project \cite{DESDR2}.

\subsection{Gold Catalog}\label{sec:y6gold}

The coadd catalog is further enhanced into a \gold catalog \cite{Y6_gold}. This is a value-added data product that includes additional columns and other ancillary data such as survey property maps, that were not included in DR2, but are used for galaxy clustering analyses for Y6 data among other applications. This catalog is the basis of the BAO sample, in particular the 2.1 version. A short summary of the main features of \gold relevant to the BAO sample is provided below.

\paragraph{A more robust and precise photometry estimate:} Flux measurements in Y6 employ a bulge plus disk model for the fit across epochs and bands, with masking of nearby objects (using the code \texttt{fitVD}, see Section 3 of \cite{2022MNRAS.509.3547H} for a description). As a change with respect to the Y3 Gold approach, the bulge and disk size ratio has been fixed in order to improve the robustness of the measurement and reduce uncertainties in the derived parameters.

\paragraph{An improved star-galaxy classifier:} The quantity \extmash measures the deviation of an object from a point-like source using 5 categories ranging from 0 (most point-like) to 4 (most extended-like). These categories are created as regions in the size (\bdft) vs signal-to-noise (\bdfsn) space of the \fitvd \footnote{\url{https://github.com/esheldon/fitvd}} quantities of the \gold photometry. In those cases where \fitvd is not available, we use \sextractor variables (see \cite{DESDR2}).

\paragraph{Additional quality flags:} The column \flagsgold is a bitmask that summarizes a collection of flags coming from the detection and measurement algorithms and at the same time adds specific features of DES images that have shown up during the years, to avoid including them in standard analyses.

\paragraph{A pixelized footprint mask with detection fraction information:} An angular mask in \healpix format containing a positive value for a given pixel if
\begin{enumerate}
    \item it has, at least, 2 exposures in each of the $griz$ bands.
    \item it covers, at least, 50\% of the \healpix combined $griz$ coverage area.
\end{enumerate}
This value is equal to the combined $griz$ coverage area, determined by a higher resolution subpixelization. In addition, each object has a \flagsfootprint value with this information as well and at the same time an assurance that the object has been indeed observed through the \nitermodel variable in $griz$.

\paragraph{An astrophysical foregrounds mask:} The \gold data set incorporates a mask that selects regions marked as having potentially problematic astrophysical foregrounds, such as bright stars from the 2MASS catalog, large nearby galaxies or extended globular clusters and dwarf spheroidals. This mask is incorporated into the angular mask used to select the BAO sample and estimate the galaxy clustering, as described in \autoref{sec:angular_mask}.

\paragraph{Survey property maps:} The \gold survey property maps are data structures that track the spatial distribution on the sky of specific observation characteristics or astrophysical measurements, which might impact the detectability of sources and their features. We use these maps (in \healpix format) to reduce the effect of systematic errors on galaxy clustering, as described in \cite{10.1093/mnras/stab2995} and detailed for Y6 data in \autoref{sec:systematics}.

\paragraph{A photometric redshift estimate:} In Y6, the fiducial photometric redshift estimate is DNF, which is described in \autoref{sec:dnf}.

The \gold catalog version used for this analysis corresponds to the internal release version 2.1, which has some minor differences with the upcoming publicly released \gold catalog (version 2.2). These differences include:
\begin{itemize}
\item In \gold version 2.1, the \fitvd photometry used in the DNF estimates is slightly modified ($\mathcal{O}$(mmag)) with respect to the photometry in the tables, corresponding to small differences in photometric corrections applied to the magnitudes.
\item In \gold version 2.1, the DNF estimates include the $Y$ band, to ensure better coverage at higher redshift. At the same time, the robustness of the measurement is more insensitive to $Y$ band survey property systematics (\autoref{sec:systematics}).
\item In \gold version 2.1, the FLAGS\_FOOTPRINT flag includes detection fraction information, which is separated in subsequent versions into footprint binary mask and survey property detection fraction mask.
\item In \gold version 2.1, additional masking on three particular tiles that had corrupt flux values was added, totaling $\sim 1.5$ square degrees.
\end{itemize}

\subsection{DNF Redshifts}\label{sec:dnf}

In order to assign galaxies to each redshift bin, we use the photo-$z$ estimate given by the Directional Neighborhood Fitting (DNF) algorithm \cite{de2016dnf}, which was trained using $grizY$\footnote{In the case of the Y3 analysis, we did not include the magnitudes in the $Y$ band, but we do in Y6.} magnitudes onto a large spectroscopic reference sample. DNF is a non-parametric method that uses a training set of galaxies with known spectroscopic redshifts to establish the relationship between the observed magnitudes and the true redshifts. The training set, compiled and validated in \cite{2018A&C....25...58G}, is described in \cite{Y6_gold}.

DNF works by fitting a linear function in the neighborhood (magnitude-color space) to the target galaxy within the training set, where the function predicts the redshift of a galaxy based on its magnitudes. The key point of DNF is that it takes into account the fact that the relationship between magnitudes and redshift may vary in different regions of the magnitude-color space. The algorithm defines a ``direction'' in the magnitude-color space to look for neighbors on the training set and thus it fits a different linear function of magnitudes for each galaxy. This allows DNF to capture more complex relationships between colors and redshift than other methods \cite{de2016dnf}.

DNF predicts the point-estimate of the photo-$z$ (called \zmean in the DES catalogs), as well as the redshift of the closest neighbor (\zmc) and the full PDF distribution\footnote{In previous DES analyses, \zmean and \zmc were referred to as {\sc Z\_MEAN} and {\sc Z\_MC}, respectively.}. The photo-$z$ estimate \zmean is computed as
\begin{equation}
    z_{\rm ph}\equiv\zmean=\sum_i\boldsymbol{a}_i\cdot\boldsymbol{m}_i,
\end{equation}
where $i$ denotes a sum over magnitudes, $\boldsymbol{a}_i$ is a parameter vector and $\boldsymbol{m}_i$ are the magnitudes in the different bands. The vector $\boldsymbol{a}_i$ is obtained by fitting the linear function using a least square regression to the set of neighbors considered.
Later, \zmean is used to assign galaxies to the redshift bins used in our BAO analysis.



\section{Sample Selection}\label{sec:baosample}

As we mentioned earlier, the same sample selection was used for both the Y1 and the Y3 BAO analyses, namely
\begin{equation}\label{eq:y3_selection}
    \begin{aligned}
    & 1.7<i-z+2(r-i)&\text{(color selection)},\\
    & 17.5<i<19+3z_{\rm ph}&\text{(flux selection)},\\
    & 0.6<z_{\rm ph}<z_{\rm ph}^{\rm max}&\text{(photo-$z$ range)},
    \end{aligned}
\end{equation}
where $r$, $i$ and $z$ are the magnitudes in the $riz$ bands, respectively; $z_{\rm ph}$ is the photometric redshift, which is given by \zmean (as defined in \autoref{sec:dnf}); and $z_{\rm ph}^{\rm max}$ is the maximum photometric redshift.
\begin{enumerate}
    \item Color selection. The color selection of \autoref{eq:y3_selection} was defined during the Y1 BAO analysis in order to select galaxies beyond $z_{\rm ph}=0.5$, following the spectral energy distribution (SED) for elliptical galaxies. Further details about this selection cut can be found in Fig. 5 of \cite{y1baosample}, where we estimated the colors of a set of SED templates as a function of redshift seen through the DES filter pass-bands. We used the same one for the Y3 analysis and we adopted it for the Y6 as well.
    \item Flux selection. In this work, we re-optimize the flux selection of \autoref{eq:y3_selection}. To do so, we leave the intercept on the y-axis and the slope as free parameters, i.e.,
    \begin{equation}\label{eq:i_selection_ab}
        17.5<i<a+bz_{\rm ph}.
    \end{equation}
    \item Photo-$z$ range. The maximum redshift, $z_{\rm ph}^{\rm max}$, was 1.0 for the Y1 and 1.1 for the Y3. For the Y6, it will be set to 1.2.
\end{enumerate}

Besides these selection cuts, by default we apply the following quality cuts to the Y6 Gold catalog\footnote{These quality cuts were eventually re-optimized at a later stage during the analysis, as described in \autoref{sec:improvement_quality}.}:
\begin{equation}
    \begin{aligned}
        \extmash \geq 3,\\
        \flagsgold = 0,\\
        \nimages[{\rm GRIZY}] > 1.
    \end{aligned}
    \label{eq:quality_cuts_y6}
\end{equation}
\extmash and \flagsgold were already defined in \autoref{sec:y6gold}. \nimages[GRIZY] is the number of $grizY$ band exposures at the object location (from the \healpix map). All these flags are described in more detail in \cite{Y6_gold}.

\subsection{Angular Mask}\label{sec:angular_mask}

\begin{figure}
    \centering
    \includegraphics[width=\linewidth]{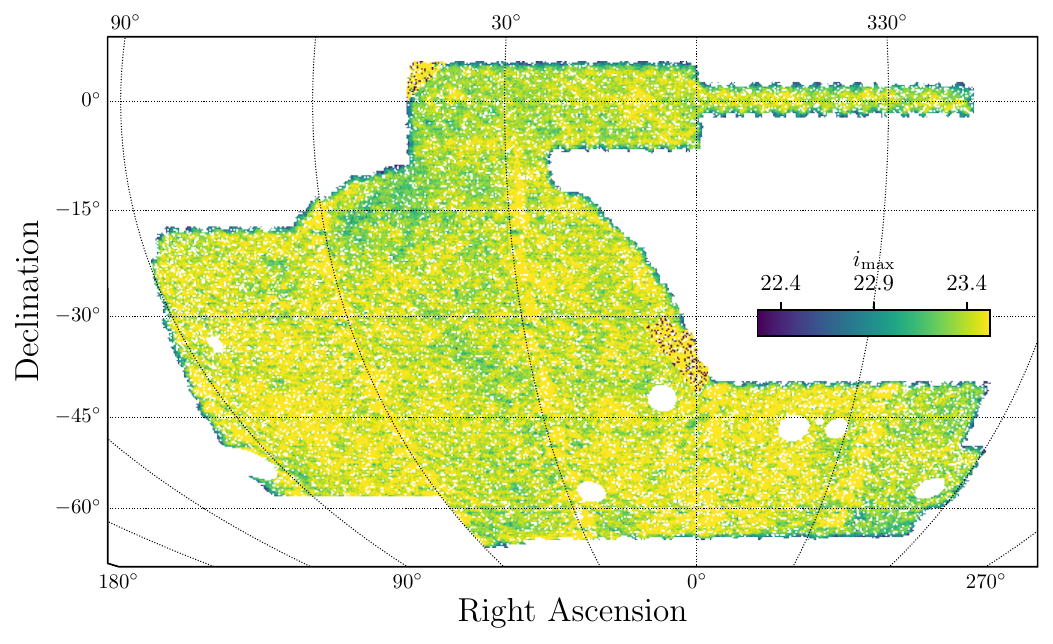}
    \caption{Footprint for the DES Y6 data. Each pixel is colored as a function of its depth in the $i$ band. The total area of the footprint considering the detection fraction of each pixel is 4,374.20 deg${}^2$.}
    \label{fig:angular_mask_Y6}
\end{figure}

The angular mask is constructed similarly to the one we used for the Y3 analysis, see \cite{10.1093/mnras/stab2995}. In order to build it, we required:
\begin{enumerate}
    \item pixels must be in the \gold footprint (see \autoref{sec:y6gold}) and have an effective coverage $>80\%$.
    \item pixels must not be affected by foreground sources, like regions around bright stars or extended galaxies.
    \item pixels must have a $10\sigma$ depth in $i$ greater than 22.5.
\end{enumerate}
The resultant footprint is shown in \autoref{fig:angular_mask_Y6}, showing the $i$-band depth in each pixel. It is worth mentioning that this is the parent mask, not the final mask, from which our BAO sample is made.

\begin{figure}
    \centering
    \includegraphics[width=\linewidth]{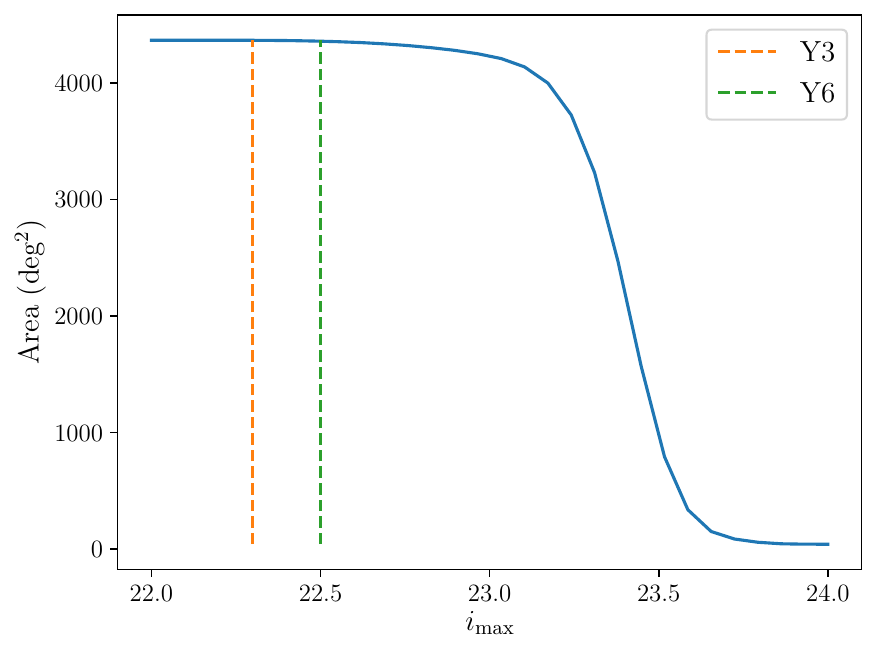}
    \caption{Area of the Y6 footprint mask as a function of the $i$-magnitude limit (blue solid line). The orange dashed line indicates the limit for the Y3 analysis ($i_{\rm max}=22.3$), whereas the green dashed line represents the same but for the Y6 ($i_{\rm max}=22.5$). We find that we barely lose any area after applying the $i=22.5$ cut (we still have 4,357.01 deg${}^2$ from the total of 4,374.20 deg${}^2$).}
    \label{fig:area_vs_i_max_linear}
\end{figure}

Different to the previous release, for the Y6 analysis we aim to optimize the sample selection as a function of the $i$-magnitude cut, see \autoref{eq:i_selection_ab}. Therefore, it is necessary to carefully account for the depth maps related to our footprint. Each pixel plotted in \autoref{fig:angular_mask_Y6} reaches a different depth in the $i$ band, which 
effectively limits the area of the mask as a function of the $i$-magnitude cut that we set, $i_{\rm max}$. In \autoref{fig:area_vs_i_max_linear} we show the area of the angular mask as a function of this $i_{\rm max}$. The deeper we want our sample to be, the more area of the full footprint we need to remove. The orange dashed line shown in this figure corresponds to the Y3 $i$-magnitude limit, which was simply $i_{\rm max}=19+3 z_{\rm ph}^{\rm max}=19+3\times 1.1=22.3$ (directly computed using $z_{\rm ph}^{\rm max}=1.1$). The green dashed line corresponds to the $i$-magnitude limit chosen for the Y6, which will be set to 22.5. The area of such mask is 4,357.01 deg${}^2$, which can be compared with the total area of the original angular mask, which is 4,374.20 deg${}^2$: we conclude that we barely lose any area by setting the $i=22.5$ limit. The final version of the Y6 BAO angular mask will have a slightly smaller area, 4,273.42 deg${}^2$ (see \autoref{sec:improvement_quality} for further details).

Since for the Y6 we have better data quality than for the Y3, i.e., less noisy magnitude estimations (because of the longer exposure time), we can afford to go deeper in magnitude (and also in redshift). However, we cannot arbitrarily go to higher magnitudes: first, because we would lose area; second, because photo-$z$ precision worsens as magnitude increases in the faint limit; and third, galaxies would be more affected by observational systematics. In order to mitigate the impact of redshift uncertainties and inaccuracies, we choose to limit our sample to
\begin{equation}\label{eq:i_22.5}
    i<22.5.
\end{equation}
The reason to choose this particular value is that we do not have reference spectra for fainter galaxies, i.e., we would not be able to trust and/or validate the photo-$z$ of a fainter galaxy sample. In fact, to calibrate the photometric redshifts in \autoref{sec:photoz} we use VIPERS, which is a complete spectroscopic sample above redshift 0.5, but only up to $i<22.5$ (see \cite{scodeggio2018vimos} for further details).

\subsection{Optimization of the Selection Cuts}\label{sec:optimization}


\subsubsection{Forecast Method}\label{fisher_forecast}

The forecasting method we use is based upon the methodology developed in \cite{seo2007improved}. Following \cite{seo2007improved, tegmark1997measuring} and assuming the likelihood function of the band powers of the galaxy power spectrum to be Gaussian, the Fisher matrix can be approximated as 
\begin{align}
    F_{ij}& =\int_{\boldsymbol{k}_{\rm min}}^{\boldsymbol{k}_{\rm max}}\frac{d^3\boldsymbol{k}}{2(2\pi)^3}V_{\rm eff}(\boldsymbol{k})\frac{\partial\log P_G(\boldsymbol{k})}{\partial p_i}\frac{\partial\log P_G(\boldsymbol{k})}{\partial p_j}\nonumber\\
    & =\int_{-1}^1d\mu\int_{k_{\rm min}}^{k_{\rm max}}\frac{2\pi k^2dk}{2(2\pi)^3} V_{\rm eff}(k,\mu) \nonumber \\
    & \qquad \times
  \frac{\partial\log P_G(k,\mu)}{\partial p_i}\frac{\partial\log P_G(k,\mu)}{\partial p_j}. 
    \label{eq:fisher_matrix}
\end{align}
$P_G(\boldsymbol{k})$ is the observed galaxy power spectrum at $\boldsymbol{k}$, $\mu$ is the cosine of the angle of $\boldsymbol{k}$ with respect to the line of sight (LOS), $p_i$ are the cosmological parameters to be constrained and $V_{\rm eff}$ is the effective volume of the survey, given by
\begin{gather}
    V_{\rm eff}(k,\mu)=\int d^3\boldsymbol{r}\left[\frac{n_{\rm gal}(\boldsymbol{r})P_G(k,\mu)}{n_{\rm gal}(\boldsymbol{r})P_G(k,\mu)+1}\right]^2\nonumber\\
    =\left[\frac{(1+\beta\mu^2)^2P(k)}{(1+\beta\mu^2)^2P(k)+n_{\rm gal}^{-1}}\right]^2V_{\rm survey}.
    \label{eq:V_eff}
\end{gather}
Here, $n_{\rm gal}(\boldsymbol{r})$ is the comoving number density of galaxies (that we assumed constant in angular position) and $\beta$ is the linear redshift-space-distortion parameter. Also, $V_{\rm survey}$ is given by
\begin{equation}
    V_{\rm survey}=\frac{4\pi}{3}f_{\rm sky}\left[\chi(z_{\rm max})^3-\chi(z_{\rm min})^3\right].
\end{equation}

This Fisher matrix can then be approximated based on how well we can center the location of the baryonic peak, i.e., the sound horizon scale $s_o$ at the drag epoch when observed in the reference cosmology. Following \cite{seo2007improved}, the fractional error on the location of the peak can be written as
\begin{gather}
    \sigma_{\log s_o}=\frac{\sigma_{s_o}}{s_o}=\sqrt{F^{-1}_{\log s_o}}
    =\left\{V_{\rm survey}A_0^2\int_{k_{\rm min}}^{k_{\rm max}} dk\right.\nonumber\\
    \left.\times \frac{k^2\exp\left(-2(k\Sigma_{\rm Silk})^{1.4}\right)\exp\left(-2k^2\Sigma_{\rm tot}^2\right)}{\left(\frac{P(k)}{P_{0.2}}+\frac{1}{n_{\rm gal}P_{0.2}}\right)^2}\right\}^{-1/2},
    \label{eq:fisher_matrix_new}
\end{gather}
where $A_0$ is a constant factor normalizing the baryonic power spectrum \cite{seo2007improved}, $P_{0.2}$ is the galaxy power spectrum at $k=0.2$ $h$/Mpc at the given redshift and the factors $\Sigma_{\rm Silk}$ and $\Sigma_{\rm tot}$ give the broadening of the BAO peak with a Gaussian function due to the Silk damping effect and the Lagrangian displacement, respectively. From \autoref{eq:fisher_matrix_new} it follows that the distance precision depends only on the survey volume, the number density of galaxies and the redshift of the survey. However, for photometric redshift surveys such as DES, it also depends on the width of the photo-$z$ distribution, $\Sigma_z$, since photometric redshift errors result in an exponential suppression of the power spectrum,
\begin{equation}
    P\to P\exp(-k^2\mu^2\Sigma_z^2).
\end{equation}

$\sigma_{\log s_o}$ is equivalent to the fractional error on the distance estimation when the physical location of the peak is well known from the CMB \cite{seo2007improved}. We compute it for each redshift bin and then combine these as
\begin{equation}
   \sigma_{\rm BAO}=\left[\sum_{\rm zbin}\frac{1}{(\sigma_{\log s_o}^{\rm zbin})^2}\right]^{-1/2}.
\end{equation}
In order to run the forecasts, we need:
\begin{itemize}
    \item $z_{\rm min}$ and $z_{\rm max}$.
    \item the area of the angular mask, $A_{\rm mask}$, which depends on the $i$-magnitude limit of the sample (see \autoref{fig:area_vs_i_max_linear}). 
    It allows us to compute $f_{\rm sky}$.
    \item the value of $\Sigma_z$ for each individual redshift bin (calculated using the expression given in \appendixcite{app:photoz}).
    \item the number of galaxies, $N_{\rm gal}$, in each redshift bin, from which we compute the number density as
    \begin{equation}
        n_{\rm gal}=\frac{N_{\rm gal}}{A_{\rm mask}}.
    \end{equation}
\end{itemize}


\subsubsection{Optimization Algorithm}


Here we describe the algorithm developed to optimize the sample selection. We first run the algorithm in 5-bin samples with photo-$z$ between 0.6 and 1.1 and then extend the analysis to 6-bin samples with photo-$z$ between 0.6 and 1.2.

In order to optimize the sample selection for the best BAO scale measurement, we need to include, at least, one free parameter in our sample selection. We add this freedom in the flux selection leaving the slope and the intercept on y-axis as free parameters: $17.5<i<a+bz_{\rm ph}$, as we discussed earlier (the other cuts are fixed by the survey characteristics).
Setting $a=19$ and $b=3$ corresponds to the Y3 selection, \autoref{eq:y3_selection}. As mentioned earlier, we impose an extra cut requiring $i<22.5$, \autoref{eq:i_22.5}.
We allow $a$ and $b$ to vary in the ranges
\begin{equation}\label{eq:a_and_b_limits}
    19\leq a\leq 22,\quad 1.5\leq b\leq 3.5,
\end{equation}
with 100 linearly-spaced values in each interval (for a total of 10,000 test samples), in order to search for their optimal values. The optimization algorithm works as follows:
\begin{enumerate}
    \item Select a pair of values for $a$ and $b$.
    \item Compute $i_{\rm max}=\text{min}(a+b z_{\rm ph}^{\rm max},22.5)$, where $z_{\rm ph}^{\rm max}=1.1$\footnote{As we already mentioned, we run the optimization algorithm for 5-bin samples in the redshift range $0.6<z_{\rm ph}<1.1$ first and then add another redshift bin from $1.1<z_{\rm ph}<1.2$ and run the algorithm for 6-bin samples.}. By default, we set the $i$-magnitude limit to 22.5. However, depending on the values of $a$ and $b$, for some samples $a+b z_{\rm ph}^{\rm max}<22.5$ and, therefore, we would unnecessarily lose area for them if we simply set $i_{\rm max}=22.5$. Therefore, the correct way to compute $i_{\rm max}$ for a given sample is to calculate the minimum between $a+b z_{\rm ph}^{\rm max}$ and 22.5.
    \item Remove pixels with depth in the $i$-magnitude band smaller than $i_{\rm max}$ from the angular mask. Compute the total area of the remaining pixels, taking into account the detection fraction of each of them. This step implies using a different angular mask for each sample (as a function of the value of $i_{\rm max}$).
    \item Create a galaxy sample applying the Y6 quality cuts, defined by \autoref{eq:quality_cuts_y6} and also the corresponding selection cuts to the Y6 Gold Catalog.
    \item Compute $\sigma_{68}$ and count the number of galaxies $N_{\rm gal}$ in each redshift bin for the galaxy sample created in step 4.
    \item Compute $\sigma_{\rm BAO}$ with the Fisher forecast code using the area of the angular mask (computed in step 3.), $\sigma_{68}$ and $N_{\rm gal}$ (both of them computed in step 4.), as explained in \autoref{fisher_forecast}.
\end{enumerate}
We apply this algorithm to the grid in the $(a,b)$ plane defined by \autoref{eq:a_and_b_limits} and find that the minimum value for $\sigma_{\rm BAO}$ is
\begin{equation}\label{eq:forecast_Y6opt_5bin}
    \sigma_{\rm BAO}^{\rm Y6-opt}=0.0170.
\end{equation}
The corresponding optimal parameters are $a=19.64$ and $b=2.894$, which are well within the limits of the $(a,b)$ plane we previously defined. 

As already mentioned, Y6 photometric redshifts are more accurate and therefore we can further optimize the BAO sample by increasing the photo-$z$ range, i.e., adding one more redshift bin from 1.1 to 1.2.
We run the forecasts in the same $(a,b)$ plane as before and find that the minimum value for $\sigma_{\rm BAO}$ is
\begin{equation}
    \sigma_{\rm BAO}^{\rm Y6-opt}=0.0162,
\end{equation}
which is smaller than the minimum value for the 5-bin case displayed in \autoref{eq:forecast_Y6opt_5bin}, i.e., $\sigma_{\rm BAO}$ decreases when adding one extra redshift bin, as expected. The optimal parameters found for the 6-bin case are exactly the same as the ones we found for the 5-bin one, namely
\begin{equation}\label{eq:optimal_param_6bin}
    \begin{aligned}
        & a_{\rm Y6-opt}=19.64,\\
        & b_{\rm Y6-opt}=2.894,
    \end{aligned}
\end{equation}
i.e., adding one extra redshift bin does not impact the optimal values for $a$ and $b$.
\begin{figure}
    \centering
    \includegraphics[width=1\linewidth]{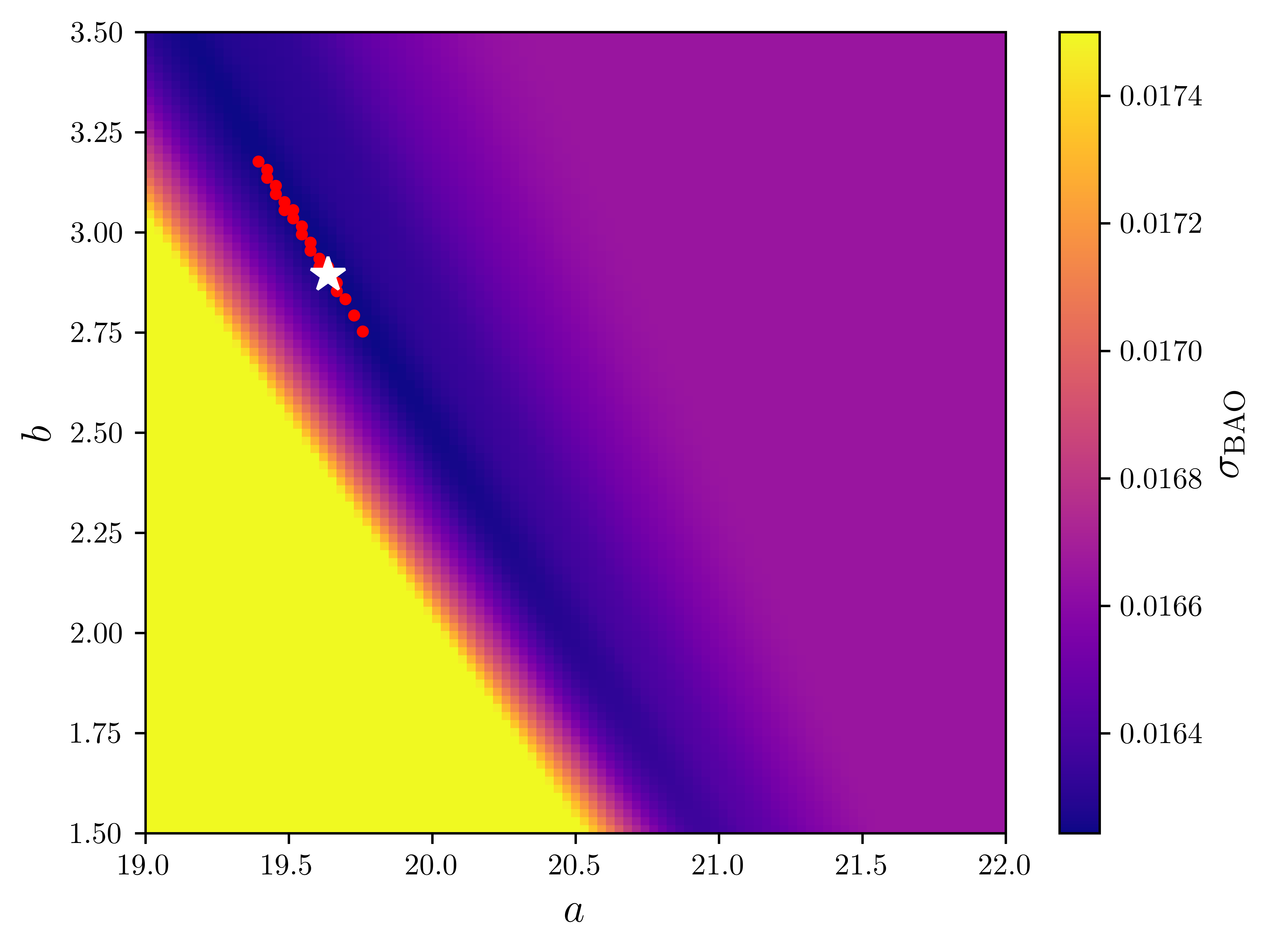}
    \caption[Heat-map of $\sigma_{\rm BAO}$ obtained for samples selected with different values of $a$ and $b$.]{Heat-map of $\sigma_{\rm BAO}$ obtained for samples selected with different values of $a$ and $b$ following \autoref{eq:i_selection_ab}. The white star represents the sample with the lowest $\sigma_{\rm BAO}$, whereas the red points correspond to the next 20 samples with lower values for this variable. The optimal sample has $a=19.64$ and $b=2.894$ and a value of $\sigma_{\rm BAO}=0.0162$.}
    \label{fig:sigma_BAO_map_6}
\end{figure}

In \autoref{fig:sigma_BAO_map_6} we show the $\sigma_{\rm BAO}$ heat-map obtained from the forecasts as a function of $a$ and $b$. The sample with the lowest $\sigma_{\rm BAO}$ is shown as a white star, but we also include the next 20 samples with the lowest values of $\sigma_{\rm BAO}$ as red points. We find that all of them lie in a diagonal-like region in which the forecasted error reaches its minimum value, i.e., all the samples in this region have, approximately, the same $\sigma_{\rm BAO}$.
We studied several properties for all these samples: the width of the photo-$z$ distribution, the total number of galaxies, the limiting magnitude and the number of photo-$z$ outliers. However, we did not find any significant difference between them: all their properties were quite similar.
Therefore, we decided to choose the values of $a$ and $b$ corresponding to the sample with the lowest $\sigma_{\rm BAO}$, i.e., the ones displayed in \autoref{eq:optimal_param_6bin}.
Therefore, the final selection of the Y6 sample is given by
\begin{equation}\label{eq:y6_selection}
    \begin{aligned}
        & 1.7<i-z+2(r-i),\\
        & 17.5<i<19.64+2.894 z_{\rm ph},\\
        & i<22.5,\\
        & 0.6<z_{\rm ph}<1.2.
    \end{aligned}
\end{equation}

In \autoref{tab:summary_fisher_forecast} we summarize the results of the forecast applied to several different samples. We include the results for the Y3 BAO sample, a Y6 sample selected using the Y3 cuts with 5 and 6 redshift bins (i.e., applying \autoref{eq:y3_selection} with $z_{\rm ph}^{\rm max}=1.1$ and 1.2, respectively) and the Y6 optimal sample with 5 and 6 redshift bins. We note that the 6-bin cases always improve with respect to the 5-bin ones. We also find that the optimal 5-bin case is already better than the Y3-selection-like 6-bin one (forecasted errors of 1.70\% and 1.76\%, respectively). The lowest value of $\sigma_{\rm BAO}$ corresponds to the 6-bin optimal Y6 sample, as expected. From these numbers, we conclude that we expect the error associated with the BAO distance measurement to be reduced by about 25\% with respect to the Y3 analysis, i.e., it goes from 2.14\% to 1.62\%, which is an important increase in precision. Part of this increase in precision is due to the higher quality of the Y6 data (2.14\% to 1.85\%), part is due to the optimization of the selection cuts (1.85\% to 1.70\%) and part is due to the increase in redshift (1.70\% to 1.62\%).
\begin{table}[]
    \centering
    \renewcommand{\arraystretch}{1.3} 
    \begin{tabular}{c|c}
        \toprule\toprule
        Case  & $\sigma_{\rm BAO}$ \\\hline
        Y3 \dotfill & 0.0214 \\
        Y6-Y3sel (5 redshift bins) \dotfill & 0.0185 \\
        Y6-Y3sel (6 redshift bins) \dotfill & 0.0176 \\
        Y6-opt (5 redshift bins) \dotfill & 0.0170\\
        \textbf{Y6-opt (6 redshift bins)} \dotfill & \textbf{0.0162}\\
        \bottomrule\bottomrule
    \end{tabular}
    \caption{Summary of the results of the Fisher forecast code applied to different samples. We include the cases of running the code using the properties of the Y3 sample; those of the Y6 sample selected with the same selection cuts as in the Y3 (5 and 6 redshift bins); and also those of the optimal Y6 sample (5 and 6 redshift bins). As expected, the most precise sample to measure the BAO feature is the optimal one with 6 redshift bins (highlighted case).}
    \label{tab:summary_fisher_forecast}
\end{table}

\subsection{Improvement of the Quality Cuts}\label{sec:improvement_quality}

In order to remove objects with large magnitude errors, additional magnitude cuts were imposed to the selection, which in consequence produced poor photo-$z$ estimates. On average, these cuts ensure that galaxies have a signal-to-noise ratio greater than 3 in $grz$ bands: 
\begin{equation}
    \begin{aligned}
        & g<25.5,\\
        & r<25,\\
        & z<24.
    \end{aligned}
\end{equation}
The star-galaxy separator was modified from the original $\textsc{EXT\_MASH}\geq 3$, see \autoref{eq:quality_cuts_y6}, to $\textsc{EXT\_MASH}=4$, which proved to better remove the remaining stellar contamination. The Y6 BAO mask was also slightly modified in order to remove regions with globular clusters and image artifacts, as detailed \autoref{sec:mask_systematics}. The final mask has an area of 4,273.42 deg${}^2$. Around $5\%$ of the galaxies of the Y6 optimal sample were removed with the combined effect of applying the new quality cuts and the modified angular mask. 

In \autoref{tab:y6_sample_properties} we display the properties of the final version of the Y6 BAO sample. We find that in the Y6 analysis we have, approximately, doubled the number of galaxies in our sample with respect to the Y3 \cite{y3baosample}. The main reason for this is the new $i$-magnitude limit of $i<19.64+2.894z_{\rm ph}$ vs $i<19+3z_{\rm ph}$ in Y3, which allows us to go deeper in every redshift bin, e.g., $i<21.67$ in the first redshift bin for Y6 vs $i<21.1$ in the Y3.
\begin{table}
    \setlength{\tabcolsep}{4pt} 
    \centering
    \begin{tabular}{ccc}
        \toprule\toprule
        Bin & $N_{\rm gal}$ & $\sigma_{68}$ \\\hline
        $0.6<z_{\rm ph}<0.7$ \ & 2,854,542 \ & 0.0232 \\
        $0.7<z_{\rm ph}<0.8$ \ & 3,266,097 \ & 0.0254 \\
        $0.8<z_{\rm ph}<0.9$ \ & 3,898,672 \ & 0.0292 \\
        $0.9<z_{\rm ph}<1.0$ \ & 3,404,744 \ & 0.0358 \\
        $1.0<z_{\rm ph}<1.1$ \ & 1,752,169 \ & 0.0403 \\
        $1.1<z_{\rm ph}<1.2$ \ &   761,332 \ & 0.0415 \\
        \bottomrule\bottomrule
    \end{tabular}
    \caption{Main properties of the Y6 BAO sample as a function of redshift: number of galaxies and dispersion on the photo-$z$, computed using \zmc and \zmean as described in \appendixcite{app:photoz}. The sample covers 4,273.42 deg${}^2$, with a total of 15,937,556 galaxies. The effective redshift of the sample is $z_{\rm eff}=0.867$, as computed in \cite{Y6_BAO_measurement}.}
    \label{tab:y6_sample_properties}
\end{table}

\section{Redshift Calibration}\label{sec:photoz}


In this section, we validate the redshift distributions of our BAO sample. Even though we could just use the VIPERS \zspec values for this (spectroscopic complete sample within our redshift/magnitude selection), we supplement them with the redshift distributions estimated using clustering redshifts (WZ). Both these methods produce somewhat noisy redshift distributions, which is the reason why we use the smoother DNF redshift distributions as templates and shift and stretch them with respect to these two as our fiducial choice (which we label as ``fiducial'' throughout this paper). This section is divided into three subsections: in \autoref{sec:photo-z_validation}, we perform a direct calibration of our photometric redshifts using VIPERS; in \autoref{sec:clustering-z}, we estimate the redshift distributions using the WZ technique; and in \autoref{sec:shift_stretch_algorithm}, we describe the algorithm developed to shift and stretch the DNF redshift distributions to make them match the properties of VIPERS \zspec and WZ. 

\begin{figure*}
    \centering
    \includegraphics[width=0.9\linewidth]{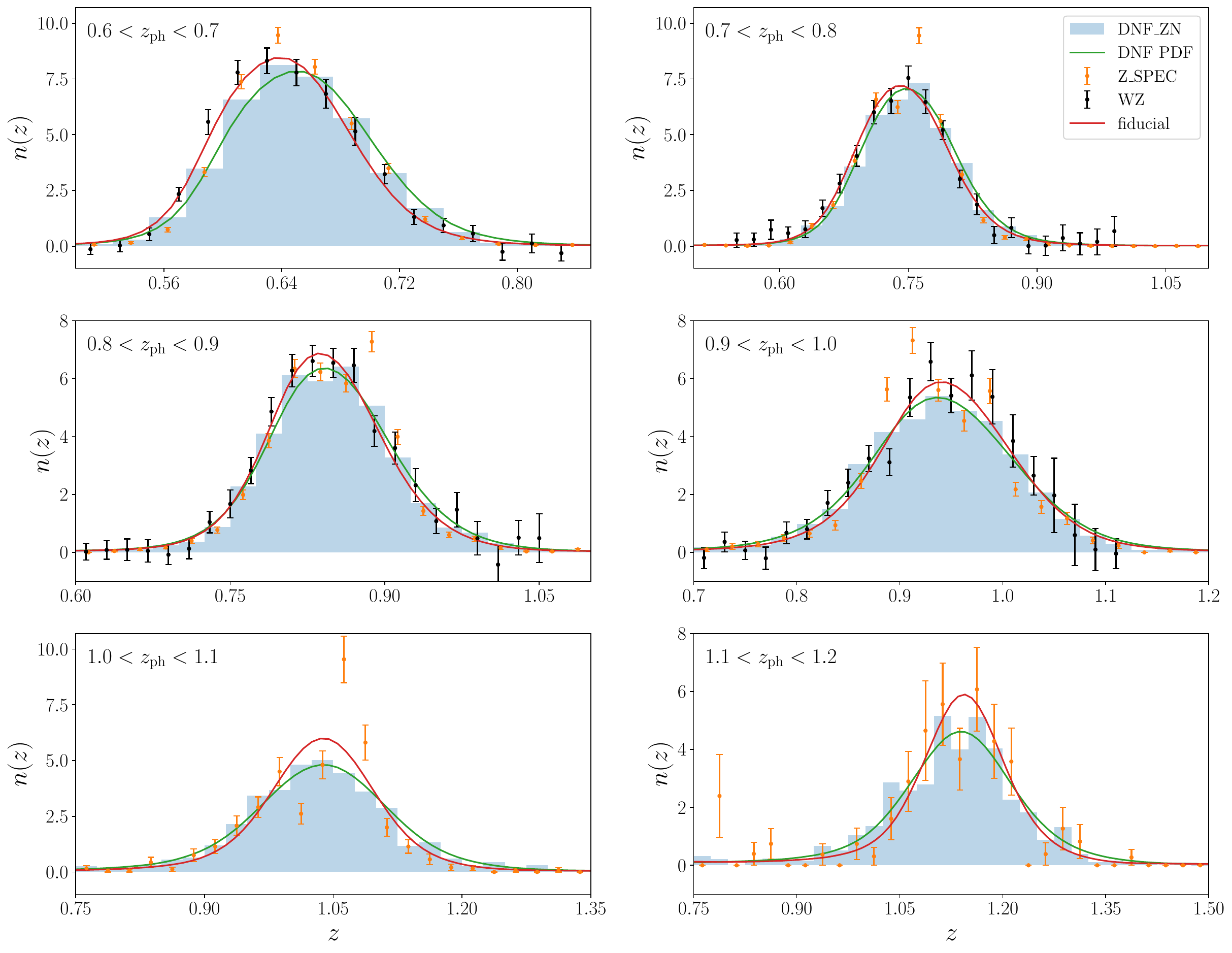}
    \caption{Redshift distributions of the Y6 BAO analysis. In the case of DNF, we show both \zmc and the stacking of DNF PDF (blue histograms and green lines, respectively). We also include the distributions of VIPERS \zspec (orange points with error-bars), WZ (black points with error-bars) and the fiducial choice (red lines), which corresponds to the redshift distributions of DNF PDF but shifted and stretched with respect to WZ in the first 4 redshift bins and with respect to VIPERS \zspec in the last 2, following the methodology described in \autoref{sec:shift_stretch_algorithm}. The mean and width of all these redshift distributions (computed using the expressions given in \appendixcite{app:photoz}) are displayed in \autoref{tab:z_average_w_68_all} and plotted in \autoref{fig:mean_z_W_68_all}.}
    \label{fig:nz_lots}
\end{figure*}

In \autoref{fig:nz_lots} we show the redshift distributions for all the different methods we just mentioned. DNF provides two alternative estimations of the redshift distributions: $n(\zmc)$ and PDF. The first one is obtained as histograms of \zmc in redshift bins defined by \zmean, whereas the second one is obtained as the stacking of individual galaxy PDFs \cite{de2016dnf}. These two are shown in \autoref{fig:nz_lots} as blue histograms and green lines, respectively. We find that the distributions of \zmc are quite smooth, being the last redshift bin the noisiest one. This is somewhat expected since the last redshift bin is the one with the lowest number density and also the one for which it is more complicated to estimate the photo-$z$ (there are fewer galaxies in the spectroscopic training sample at higher redshifts). The combination of these two effects yields to a decrease in the photo-$z$ quality at such high redshifts and also makes the redshift distributions noisier. On the other hand, we find that DNF PDF is qualitatively similar to \zmc, but smoother (since it is computed as the stacking of large amounts of individual galaxy PDFs). Besides the DNF results, in \autoref{fig:nz_lots} we also include the redshift distributions of VIPERS \zspec, WZ and the fiducial choice, which are further discussed later.

\begin{figure}
    \centering
    \includegraphics[width=\linewidth]{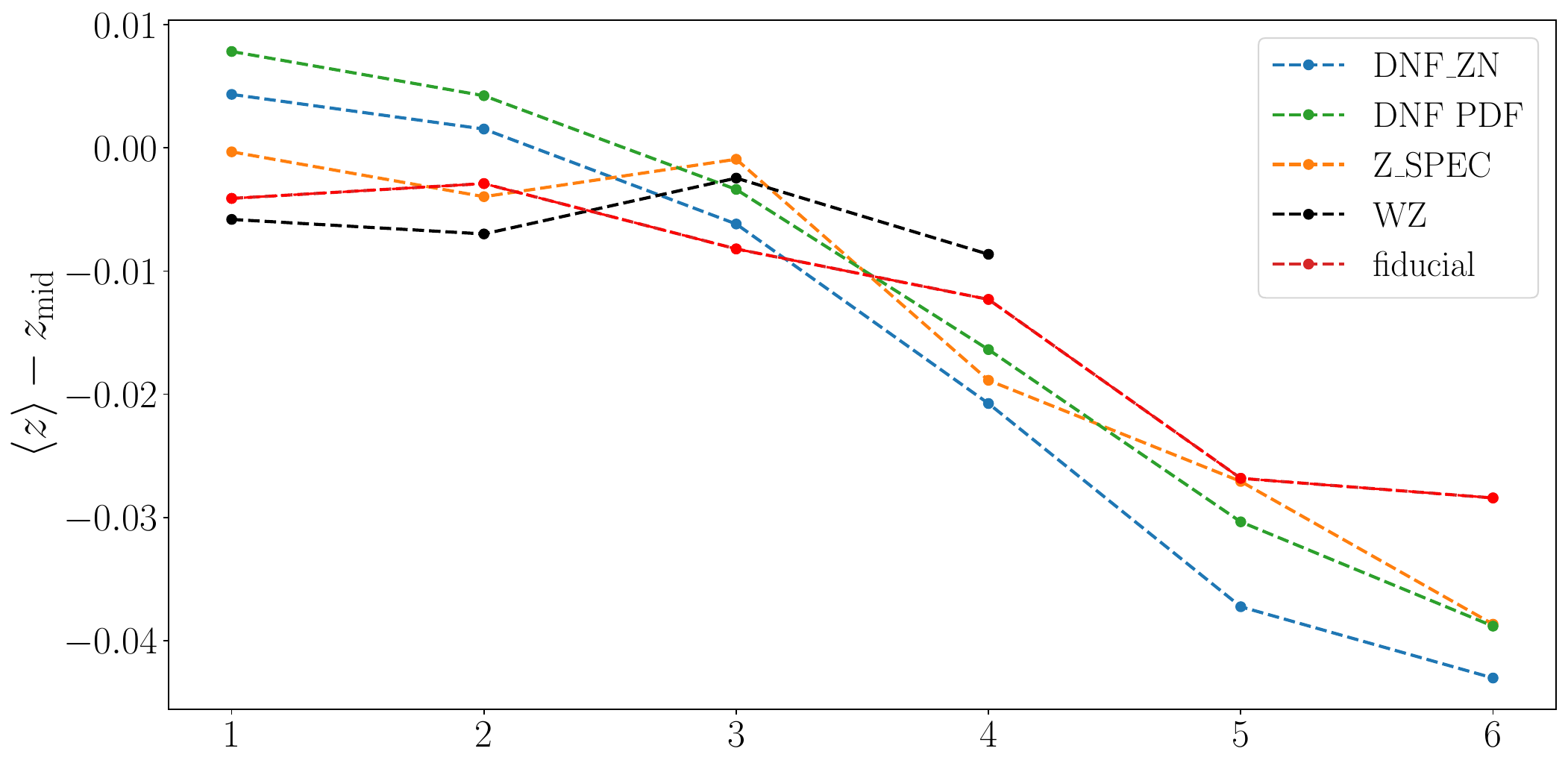}
    \includegraphics[width=\linewidth]{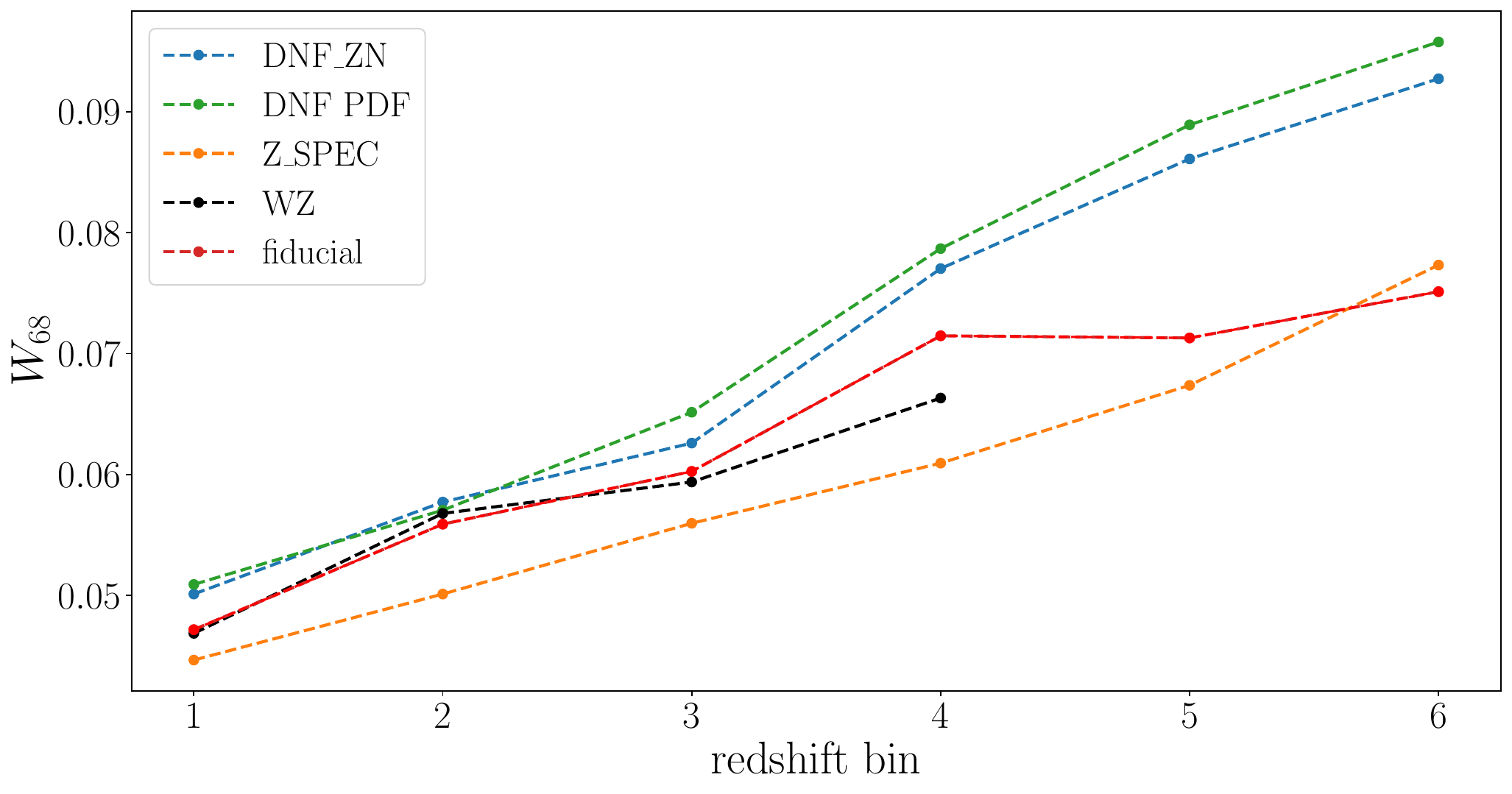}
    \caption{\textbf{Top panel:} average redshift of the different redshift distributions of the Y6 analysis. For visualization purposes, we subtracted the middle redshift, which is given by the average of the limits for each redshift bin (0.65, 0.75, 0.85, 0.95, 1.05 and 1.15, respectively). \textbf{Bottom panel:} width of the different redshift distributions of the Y6 analysis. Cases included in this plot: \zmc (blue), DNF PDF (green), \zspec (orange), WZ (black) and fiducial choice (red).}
    \label{fig:mean_z_W_68_all}
\end{figure}

In \autoref{tab:z_average_w_68_all} we show the mean and width of the different redshift distributions shown in \autoref{fig:nz_lots}. The results shown in this table were computed with the expressions given in \appendixcite{app:photoz} and are plotted in \autoref{fig:mean_z_W_68_all} for visualization purposes. We find that the properties of \zmc and DNF PDF are quite similar. Also, those of VIPERS \zspec and WZ are in very good agreement. It is also important to note that both \zmc and DNF PDF are biased and wider with respect to \zspec and WZ, which is the reason why we cannot directly use the results from DNF as the redshift distributions of our sample (and the reason why we will shift and stretch them to match the properties of VIPERS and WZ, which are closer to the underlying true redshift distributions).

\begin{table*}
    \renewcommand{\arraystretch}{1.3} 
    \setlength{\tabcolsep}{4pt} 
    \centering
    \begin{tabular}{c|ccccc|ccccc}
        \toprule\toprule
            \multirow{2}{*}{Bin} & \multicolumn{5}{c|}{$\langle z\rangle$} & \multicolumn{5}{c}{$W_{68}$} \\\cline{2-11}
            & \zmc & DNF PDF & \zspec & WZ & fiducial & \zmc & DNF PDF & \zspec & WZ & fiducial \\\hline
            $0.6<z_{\rm ph}<0.7$ & {0.654} & {0.658} & {0.650} & {0.644} & {0.646} & {0.050} & {0.051} & {0.045} & {0.047} & {0.047} \\
            $0.7<z_{\rm ph}<0.8$ & {0.752} & {0.754} & {0.746} & {0.743} & {0.747} & {0.058} & {0.057} & {0.050} & {0.057} & {0.056} \\
            $0.8<z_{\rm ph}<0.9$ & {0.844} & {0.847} & {0.849} & {0.848} & {0.842} & {0.063} & {0.065} & {0.056} & {0.059} & {0.060} \\
            $0.9<z_{\rm ph}<1.0$ & {0.929} & {0.934} & {0.931} & {0.941} & {0.938} & {0.077} & {0.079} & {0.061} & {0.066} & {0.071} \\
            $1.0<z_{\rm ph}<1.1$ & {1.013} & {1.020} & {1.023} & {---}   & {1.023} & {0.086} & {0.089} & {0.067} & {---}   & {0.071} \\
            $1.1<z_{\rm ph}<1.2$ & {1.107} & {1.111} & {1.111} & {---}   & {1.122} & {0.093} & {0.096} & {0.077} & {---}   & {0.075} \\
        \bottomrule\bottomrule
    \end{tabular}
    \caption{Mean and width of the redshift distributions, computed using the expressions given in \appendixcite{app:photoz}, for \zmc, DNF PDF, VIPERS \zspec, WZ and the fiducial choice.}
    \label{tab:z_average_w_68_all}
\end{table*}

\subsection{Direct Calibration with VIPERS Spectroscopic Redshifts}\label{sec:photo-z_validation}

We calibrate our photo-$z$ using VIPERS, similarly to what we did during the Y3 BAO analysis \cite{y3baosample}. VIPERS is a complete spectroscopic sample for redshifts above 0.5 and up to $i=22.5$, see \cite{scodeggio2018vimos} (the same $i$-magnitude limit we set for our BAO sample), with an overlapping area of 16.3 deg${}^2$ with DES. Therefore, the distributions of the spectroscopic redshifts of VIPERS in redshift bins defined by \zmean provide a direct estimation of the true redshift distributions of our BAO sample\footnote{What we actually use is not the complete VIPERS sample, but those galaxies of VIPERS that are also part of the BAO sample.} (this is explicitly shown in \appendixcite{app:redshift_calibration}). These distributions are plotted as orange points with error-bars in \autoref{fig:nz_lots}. 

One caveat is that, due to the small overlap between VIPERS and DES, the \zspec distributions are noisy, particularly in the last two redshift bins. Also, as we mentioned earlier, the distributions of \zspec are narrower than those from DNF (see \autoref{tab:z_average_w_68_all}). For these two reasons, we use \zspec to shift and stretch the DNF redshift distributions for redshifts above 1.0. For redshifts below 1.0, we use clustering redshift, which we describe next.



\subsection{Clustering Redshift (WZ)}\label{sec:clustering-z}

There are alternative ways to estimate the redshift distributions of our BAO sample, such as the clustering redshift technique \cite{Newman:2008}. Clustering redshifts make use of the fact that galaxies with unknown redshifts reside in the same structures as galaxies that have known redshifts. Thus, spatial cross-correlations can be used to estimate the redshift distribution of the sample with unknown redshifts. The modern approach of using this data to obtain a precise estimate of a redshift distribution can be traced back to \cite{Newman:2008}. Since then, it has been implemented and further developed in the literature, see \cite{2010ApJ...721..456M,2013MNRAS.433.2857M,2013arXiv1303.4722M,2016MNRAS.463.3737C,2016MNRAS.462.1683S,10.1093/mnras/stw3033,hildebrandt2021kids,2020A&A...642A.200V} for reference.
This technique was already validated and applied to the Y3 {\sc MagLim} sample in order to calibrate its redshift distributions in \cite{cawthon2022dark} and we use the methodology choices from that work. In that case and also in this one, spectroscopic galaxies from BOSS \cite{reid2016sdss} and its extension, eBOSS \cite{2020MNRAS.498.2354R,raichoor2021completed}, are used to cross-correlate with our sample. These samples overlap about 15\% of the DES footprint.

In \autoref{fig:nz_lots} we show the redshift distributions for WZ (black points with error-bars). Because of the lack of spectroscopic galaxies in the redshift range $1.0<z<1.2$, it was not possible to estimate them for the last two bins\footnote{We do actually have SDSS galaxies in the redshift range $1.0<z<1.1$. However, because of the tails of the distribution, it was not possible to cover the whole redshift range when computing the redshift distribution in that bin.}. We find that WZ is consistent with \zspec, see the results displayed in \autoref{tab:z_average_w_68_all}. Since we have two independent determinations of the redshift distributions that agree, i.e., VIPERS \zspec and WZ, we consider them as validated.

As in the case of \zspec, WZ is also somewhat noisy. To address the problem of the noisy nature of WZ and VIPERS \zspec, for the fiducial analysis of the data we decided to use a modified version of the DNF redshift distributions as our default choice: shifted and stretched to match the properties of WZ in the first 4 redshift bins and \zspec in the last 2. Therefore, DNF $n(z)$\footnote{We can use either \zmc or DNF PDF as templates, but we decided to use the latter as our fiducial. The reason is that both have similar properties (see the results displayed in \autoref{tab:z_average_w_68_all} and/or \autoref{fig:mean_z_W_68_all}), but DNF PDF has a much smoother shape (see \autoref{fig:nz_lots}).} are used as templates for the shape of our fiducial redshift distributions.

\subsection{Shift and Stretch Algorithm}\label{sec:shift_stretch_algorithm}

Here we describe the algorithm developed to perform the shift and stretch of \zmc and DNF PDF.
We run the shift and stretch of a given $n(z)$ in two different steps:
\begin{enumerate}
    \item Shift of the original $n(z)$. The shifted redshift distribution is, simply, given by
    \begin{equation}
        n_{\rm shifted}(z,\Delta z)=n(z-\Delta z).
    \end{equation}
    \item Stretch of the shifted $n(z)$. The shifted and stretched redshift distribution is given by
    \begin{gather}
        n_{\rm 2-param}(z,\Delta z,\sigma_z)\nonumber\\
        =n_{\rm shifted}(\sigma_z(z-\langle z\rangle_{\Delta z})+\langle z\rangle_{\Delta z},\Delta z),
    \end{gather}
    where
    \begin{equation}
        \langle z\rangle_{\Delta z}\equiv\int dz\hspace{0.05cm}n_{\rm shifted}(z,\Delta z).
    \end{equation}
\end{enumerate}
We, then, compute the best fit parameters $\Delta z$ and $\sigma_z$ by minimizing
\begin{equation}
    \chi^2(\Delta z,\sigma_z)=\sum_i\left[\frac{n_{\rm 2-param}(z_i,\Delta z,\sigma_z)-n_{\rm ref}(z_i)}{\Delta n_{\rm ref}(z_i)}\right]^2.
\end{equation}
Since the redshift distributions we shift and stretch are either \zmc or DNF PDF, we neglect their contribution to the denominator of the previous expression, since their shot-noise is much smaller than that of the reference redshift distribution, which is either VIPERS or WZ. This methodology is similar to the one used for the DES Y3 3$\times$2pt analysis, see \cite{PhysRevD.105.023520}. In that context, these shift and stretch parameters appear because of our uncertainty in the photo-$z$ and this implementation is particularly useful since it allows us to fit them when running the 3$\times$2pt chains.

Using the algorithm we just described, we perform the shift and stretch of DNF PDF with respect to WZ in the first 4 redshift bins and with respect to \zspec in the last 2. The resulting redshift distributions are shown as red lines in \autoref{fig:nz_lots}, where we can check that they are as smooth as the original DNF PDF, but shifted and stretched. From the results displayed in \autoref{tab:z_average_w_68_all}, we find that the fiducial choice has a similar mean and width to those of WZ in the first 4 redshift bins and to those of \zspec in the last 2, as intended. These are the redshift distributions used to generate the BAO template to run the BAO fits on the data in \cite{Y6_BAO_measurement}.

For the projected correlation function (PCF) estimator that we use to measure the BAO in \cite{Y6_BAO_measurement}, the shift and stretch algorithm is adapted to work in 22 logarithmic redshift bins, instead of the fiducial 6 redshift bins. This is described in \appendixcite{app:calibration_pcf}.

    



\section{Correcting for Observational Systematics}\label{sec:systematics} 

In order to reduce the impact of observational systematics, we apply two complementary strategies: we first mask pixels with potentially problematic values and, next, we compute correcting weights that are applied to our galaxy sample. Both steps rely on survey property (SP) maps, which are pixel maps that keep track of spatial variations of the different systematic effects concerning the imaging of the data. Among these SP maps, we consider effects such as the seeing (FWHM) and the limiting magnitude, but also astrophysical foregrounds, such as contamination from the stellar density and galactic dust extinction.

\subsection{Masking Observational Systematics}\label{sec:mask_systematics}


As a first step to mitigate the impact of observational systematic effects, we apply an angular mask to our galaxy sample which removes potentially problematic regions of the footprint. This mask is applied on top of the Y6 angular mask we described in \autoref{sec:angular_mask} (and after running the optimization of the sample). The procedure is similar, though less strict, to the one used for the DES Y6 lens galaxy samples \cite{Y6_LSS_sys}: 
\begin{itemize}
    \item We start with the baseline mask that defines our footprint at a \healpix resolution of $N_{\rm side}$ = 4096 ($\sim$0.74 arcmin$^2$) using the criteria detailed in \autoref{sec:angular_mask}.
    \item Within this footprint, we mask regions that have image artifacts. From visual inspection, it was found that extreme values of the \emph{mean surface brightness} per pixel (SB\_MEAN) identified regions with imaging artifacts, such as the wings of bright stars. In particular, the most extreme image artifacts correspond to pixels of the SB\_MEAN quantity with values higher than 99.99\% of them, so we mask those pixels out.
    We do this separately for all SB\_MEAN maps in \emph{griz} bands. This cut removes $\sim0.02\%$ of the area (see \cite{Y6_mask,Y6_LSS_sys} for more details). 
    \item We also mask areas with an excess of diffuse emission due to galactic cirrus. We use a convolutional neural network to estimate the probability of galactic cirrus being present in a given pixel, which gives us the \emph{mean nebulosity prob} quantity, NEB\_MEAN. We use visual inspection of this variable to define a threshold of NEB\_MEAN $> 0.5$ (more details in \cite{Y6_LSS_sys,Y6_mask}). This cut is applied for each photometric band separately (\emph{griz}). 
    \item We use the intersection of these masks with a foreground mask that excludes pixels with globular clusters. 
\end{itemize}

Finally, we create a \emph{joint mask} that includes the three different cuts described above. The area of the resulting mask is 4,273.42 deg$^2$. In \autoref{tab:mask_area} we detail the area removed by each cut with respect to the baseline mask. Note that the fraction of area removed on the final joint mask is not exactly equal to the sum of the areas removed by the individual cuts, since the cirrus maps are correlated. 

\begin{table}[]
    \centering
    \begin{tabular}{ccc}
        \toprule\toprule
        Mask & Area [deg$^2$] & Removed area [\%] \\ \hline
        Baseline & 4,357.01 & - \\ 
        SB\_MEAN $> 99.99\%$ & 4,355.98 & 0.024 \\ 
        NEB\_MEAN $> 0.5$ & 4,277.24 & 1.831 \\ 
        Globular clusters & 4,354.15 & 0.066 \\
        Joint mask & 4,273.42 & 1.919 \\ 
        \bottomrule\bottomrule
    \end{tabular}
    \caption{Area removed by each of the systematic cuts with respect to the baseline mask. The final joint mask is the definitive Y6 BAO sample area. }
    \label{tab:mask_area}
\end{table}

\subsection{Galaxy Weights} \label{sec:weights}
Even after applying the quality cuts both on the sample definition and on the angular mask, there are observational effects that still may induce non-cosmological clustering signal on the galaxy density field. This is due to the variation of the observing conditions, such as seeing or sky brightness and to other aspects of the survey strategy, such as exposure time and airmass, during the period of observations. Astrophysical foregrounds, e.g., the stellar density (see \autoref{sec:stellardens}) or galactic dust, are also sources of systematic error on the clustering signal. 

The different sources of observational systematics that we consider are characterized by \healpix maps of $N_{\rm side}$ = 4096, which we refer to as survey property maps, or SP maps. Here, we detail the list of SP maps considered as our fiducial set of contamination templates (more information can be found in \cite{Y6_gold}) and \cite{Y6_LSS_sys}): 
\begin{itemize}
    \item AIRMASS (\emph{grizY}): \emph{mangle} weighted mean value of the secant of the zenith angle.
    \item FWHM (\emph{grizY}): \emph{mangle} weighted mean value of the FWHM of the 2D elliptical Moffat function that fits best the PSF model from PSFEx.
    \item SKYSIGMA (\emph{grizY}): \emph{mangle} weighted mean value of standard deviation on the sky brightness.
    \item MAGLIM (\emph{grizY}): \emph{mangle} weighted mean value of the $10\sigma$ magnitude limit in 2 arcsec aperture diameter estimated by \emph{mangle}.
    \item SFD98: $E(B-V)$ interstellar extinction map estimated from a map of dust IR emission \cite{sfd98}.
    \item GAIA: Gaia EDR3 map with $i>17$ cut \cite{gaia_edr3}.
    \item DIVOT\_edensity\_GAIA: approximation of the local background over-subtraction by fitting a simple empirical model to Gaia stars of different magnitudes by using large aperture fluxes \cite{Y6_gold}.
\end{itemize}
We note that this set of template maps is a subset of all Y6 available SP maps. We made this selection based on the same criterion as in Y3, by which we group together maps according to their spatial correlation and their physical meaning and we select a representative from each group (see \cite{y3baosample, monroy} and \cite{Y6_LSS_sys} for details on the criterion applied to Y6). 

Regarding the NEB\_MEAN and SB\_MEAN maps introduced in the previous section, they are highly non-Gaussian and, therefore, ill-suited for use in the standard regression-based algorithm described next. Furthermore, it is only the most extreme values of these maps that are problematic and thus we use them only to define the masks, but not to assign the systematic weights.

In order to account for and correct for these observational systematics on the clustering, we have applied the \emph{Iterative Systematics Decontamination} (ISD) method, which was also used for the Y3 BAO analysis. This method is described in detail in \cite{elvinpoole, monroy, y3baosample} and compared to other clustering systematic mitigation methods in \cite{ENET}. ISD starts from the hypothesis that true galaxy number density fluctuations do not correlate with those of observing conditions. The metric used to characterize the significance of the systematic contamination is the so-called 1D relation, which shows the relation between the observed galaxy number density as a function of the values of a given SP map. We compute the 1D relations by binning the SP map values in 10 bins defined in such a way that they cover equal areas on the final footprint (i.e., after applying the joint mask introduced in the previous section). From this point, the basics of ISD are synthesized as follows: 
\begin{itemize}
    \item Fix a threshold, $T_{\rm 1D}$, for the systematic contamination; 
    \item Obtain the 1D relation of observed galaxy number density $n_g$ with respect to each SP map and compute $\Delta\chi^2=\chi^2_{\rm null}-\chi^2_{\rm model}$, where $\chi^2_{\rm null}$ corresponds to the fit to null test (i.e., no systematic impact on $n_g$) and $\chi^2_{\rm model}$ corresponds to a linear fit; 
    \item Obtain $\Delta\chi^2$ from the same 1D relations between the $n_g$ measured on a set of 1000 lognormal mocks and the same SP maps; 
    \item Define the 1D significance of the systematic contamination as $S_{\rm 1D}=\Delta \chi^2/\Delta\chi^2_{68}$, where $\Delta \chi^2_{68}$ corresponds to the value that explains 68\% of the mocks. 
    \item For the SP map with the highest $S_{\rm 1D}$ and provided it is larger than our threshold, $T_{\rm 1D}$, we generate weights by taking the best linear fit function obtained on the corresponding 1D relation from the previous steps and evaluating its inverse on the $N_{\rm side}$ = 4096 pixels of that SP map.
    \item Apply the resulting weight map multiplicatively, pixel by pixel, to the BAO sample; 
    \item Repeat the process iteratively for the newly weighted sample. 
\end{itemize}

Once all the maps have $S_{\rm 1D}$ below $T_{\rm 1D}$ for the iteratively-weighted sample, the final result is a weight for each galaxy, which is computed as the product of all the weights derived in each iteration of the algorithm.

\begin{figure}
    \centering
    \includegraphics[width=\linewidth]{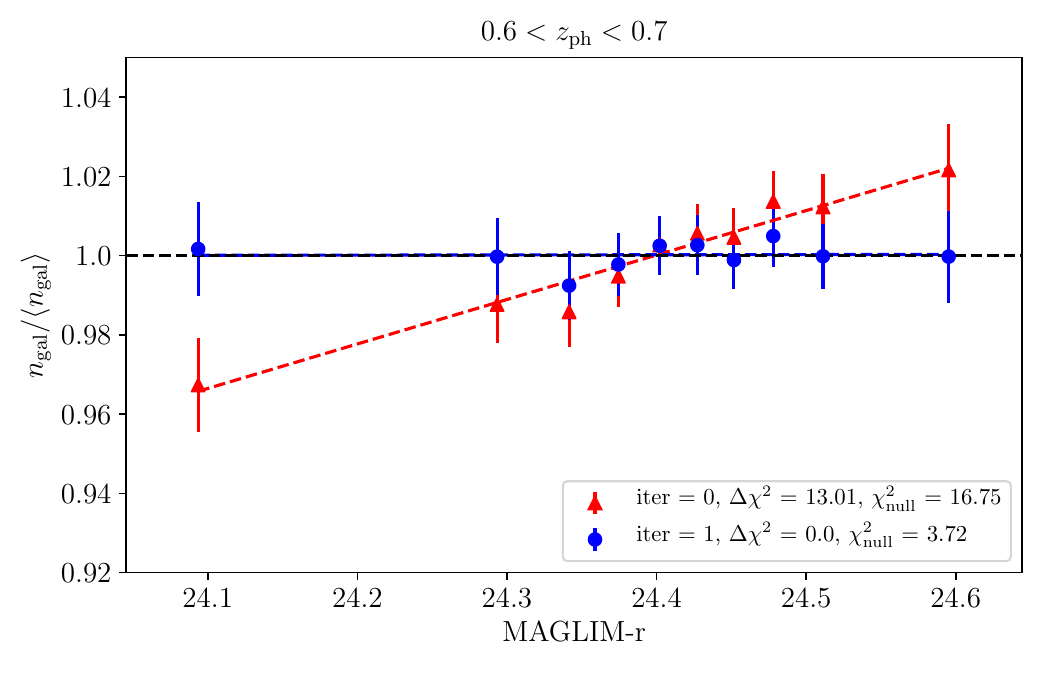}
    \caption{Example of a 1D relation from our set of fiducial SP maps, before (red triangles) and after (blue dots) being corrected using the weights obtained with ISD. Dashed lines correspond to the best linear fits in each case. }
    \label{fig:1d_plots}
\end{figure}

In \autoref{fig:1d_plots} we show, as an example, the process that one SP map undergoes after being flagged as significant by ISD. The red dashed line represents the 1D relation of the MAGLIM map in the $r$-band for the first redshift bin of the unweighted BAO sample, while the blue dashed line shows the same relation after applying correcting weights computed for this SP map. All the other 1D relations are explicitly shown in \autoref{fig:all_1ds} of \appendixcite{app:1d_plots}.

We run ISD for each redshift bin of the BAO sample independently. The significance threshold used for the Y3 BAO analysis was $T_{\rm 1D}=4$. However, validation tests on lognormal mocks showed there is no over-correction when using $T_{\rm 1D}=2$ and therefore for Y6 we decide to use this threshold, lowering the risk of under-correction biases on $w(\theta)$. 
Nevertheless, we note that the measurement of the BAO peak position is highly insensitive to the effect of the observational systematics that we consider, as tested in section VII of \cite{Y6_BAO_measurement}. This provides an additional level of protection against under- and over-corrections.

The list of SP maps found to have the most significant impact on each redshift bin is shown in \autoref{tab:spmaps_baosample}. Comparing these results with the Y3 ones (see \cite{y3baosample}), we observe a reduction in the number of SP maps we have to correct for, even if we use a stricter threshold. This is mainly due to the higher homogeneity of the survey, but also to the additional effort on masking out pixels with extreme SP values, as described in \autoref{sec:mask_systematics}, as well as the deeper Y6 data and the optimized sample selection. We note the fact that we need to correct for more maps as the redshift increases, which is expected since faint objects are more sensitive to variations in observing conditions. Among all SP maps considered, we find that those causing the most significant contaminations are the FWHM and MAGLIM maps on different photometric bands and also the two GAIA-related maps.

Finally, in order to validate the corrections provided by the systematic weights, we run a set of validation tests which we describe in \appendixcite{app:wval_tests}. 

\begin{table}[]
    \renewcommand{\arraystretch}{1.2} 
    \setlength{\tabcolsep}{4pt} 
    \centering
    \begin{tabular}{cc}
        \toprule\toprule
        Bin & SP maps used to correct for with $T_{\rm 1D} = 2$ \\ \hline
        $0.6 < z_{\rm ph} < 0.7$ & MAGLIM-r   \\
        $0.7 < z_{\rm ph} < 0.8$ & FWHM-z, GAIA, FWHM-Y \\
        $0.8 < z_{\rm ph} < 0.9$ & GAIA, FWHM-Y, FWHM-r \\
        $0.9 < z_{\rm ph} < 1.0$ & FWHM-z \\
        $1.0 < z_{\rm ph} < 1.1$ & DIVOT\_edensity\_GAIA, FWHM-z  \\
        \multirow{2}{*}{$1.1 < z_{\rm ph} < 1.2$} & DIVOT\_edensity\_GAIA, FWHM-z, \\ & FWHM-Y, MAGLIM-z \\ 
        \bottomrule\bottomrule
    \end{tabular}
    \caption{List of SP maps found to have an impact on the Y6 BAO sample at each redshift bin. }
    \label{tab:spmaps_baosample}
\end{table}

\subsection{Stellar Fraction Correction}\label{sec:stellardens}

While most observational systematics modulate the observed number density multiplicatively, $n_{\rm gal}(\theta)\rightarrow (1+f(\theta))n_{\rm gal}(\theta)$, residual stellar contamination represents an additional and undesired population in our sample, which contributes additively as $n_{\rm gal}(\theta)\rightarrow n_{\rm gal}(\theta)+n_{\rm star}(\theta)$. 
Defining the overdensity of each population as $\delta_X=n_X/\bar{n}_X-1$, with $X\in\{{\rm gal},{\rm star}\}$ and $f_{\rm star}=\bar{n}_{\rm star}/(\bar{n}_{\rm gal}+\bar{n}_{\rm star})$,
\begin{equation}
    \delta_{\rm obs}=\delta_{\rm gal}+f_{\rm star}(\delta_{\rm star}-\delta_{\rm gal}).
\end{equation}

In the limit where the density of stars is smoothly varying relative to the density of galaxies, we can approximate $\delta_{\rm star} \sim 0$. Therefore, the net effect is a suppression of galaxy fluctuations by an overall factor,
\begin{equation} \label{eq:fstarapprox}
    \delta_{\rm obs}\approx\delta_{\rm gal}(1-f_{\rm star}).
\end{equation}
This is equivalent to modifying the integral constraint and is an effect that is perfectly degenerate with linear galaxy bias, as noted in \cite{Krolewski:2019yrv}. The contribution from spatially varying contamination that is neglected in \autoref{eq:fstarapprox} is captured to first order by the standard treatment described in \autoref{sec:weights} when computing galaxy weights through the inclusion of a stellar density map from Gaia and so we only need to estimate an overall average $f_{\rm star}$ (see also \cite{LSST:2019wqx}, wherein a similar approach is applied with HSC data).

We estimate stellar contamination through the same procedure used for the Y6 lens galaxy samples \cite{Y6_LSS_sys} and refer the reader there for more detail. 
Briefly, we match our BAO sample to the public DECaLS DR9 catalog \cite{dey2019overview}, which includes forced unWISE photometry\footnote{We define matches as those objects in each catalog with $< 1$ arcsec distance between them. 99.8\% of our sample appears in DECaLS.} \cite{schlafly2019unwise}.
Because of a peak in stellar SEDs at $\sim1.6\mu$m, redshifted galaxies appear relatively brighter in the unWISE W1 band, making stars and galaxies appear in different parts of the $(r-z,z-W1)$ color-space\footnote{E.g., \cite{DESI:2022gle} used a cut in this space to remove most of the stars from the DESI LRG target sample and our approach is inspired by that work.}.
For each redshift bin, we plot the density of matched objects in the $(r-z,z-W1)$ color-space and define a piecewise linear relation that traces the trough between the peaks that are NIR-bright (galaxies) and those that are NIR-faint (stars). 

We compute $f_{\rm star}$ as the fraction of all objects on the NIR-faint side of the piecewise separation in each bin, finding 
\begin{equation}
    f_{\rm star}=[2.3\%,\ 2.7\%,\ 3.3\%,\ 2.3\%,\ 0.8\%,\ 0.7\%], 
\end{equation}
with $1\sigma$ uncertainty of roughly 0.15\%.

The resultant clustering measurements, i.e., $w(\theta)$, $C_\ell$ and $\xi_p(s_\perp)$, are corrected by a factor of $(1-f_{\rm star})^{-2}$ to account for this, though for template-based BAO measurements, such as the one we run in \cite{Y6_BAO_measurement}, this effect is negligible due to its degeneracy with other nuisance parameters.





\section{Unblinding the Clustering Measurements}\label{sec:unblinded_clustering}

As we already mentioned, in \cite{Y6_BAO_measurement} we measure the BAO distance scale using the clustering measurements from the sample optimized in this paper. The analysis is performed blind, which effectively means that the results of our measurements cannot be reported and that the clustering measurements on the data cannot be plotted until a battery of robustness tests has been passed (these tests are described in detail in \cite{Y6_BAO_measurement}). This blinding criteria is similar to the one used during the Y3 BAO analysis, see \cite{y3bao}. By the time we included this section in the paper, these tests had already been passed and we were ready to unblind the clustering measurements of our sample.

\begin{figure*}
    \centering
    \includegraphics[width=1\textwidth]{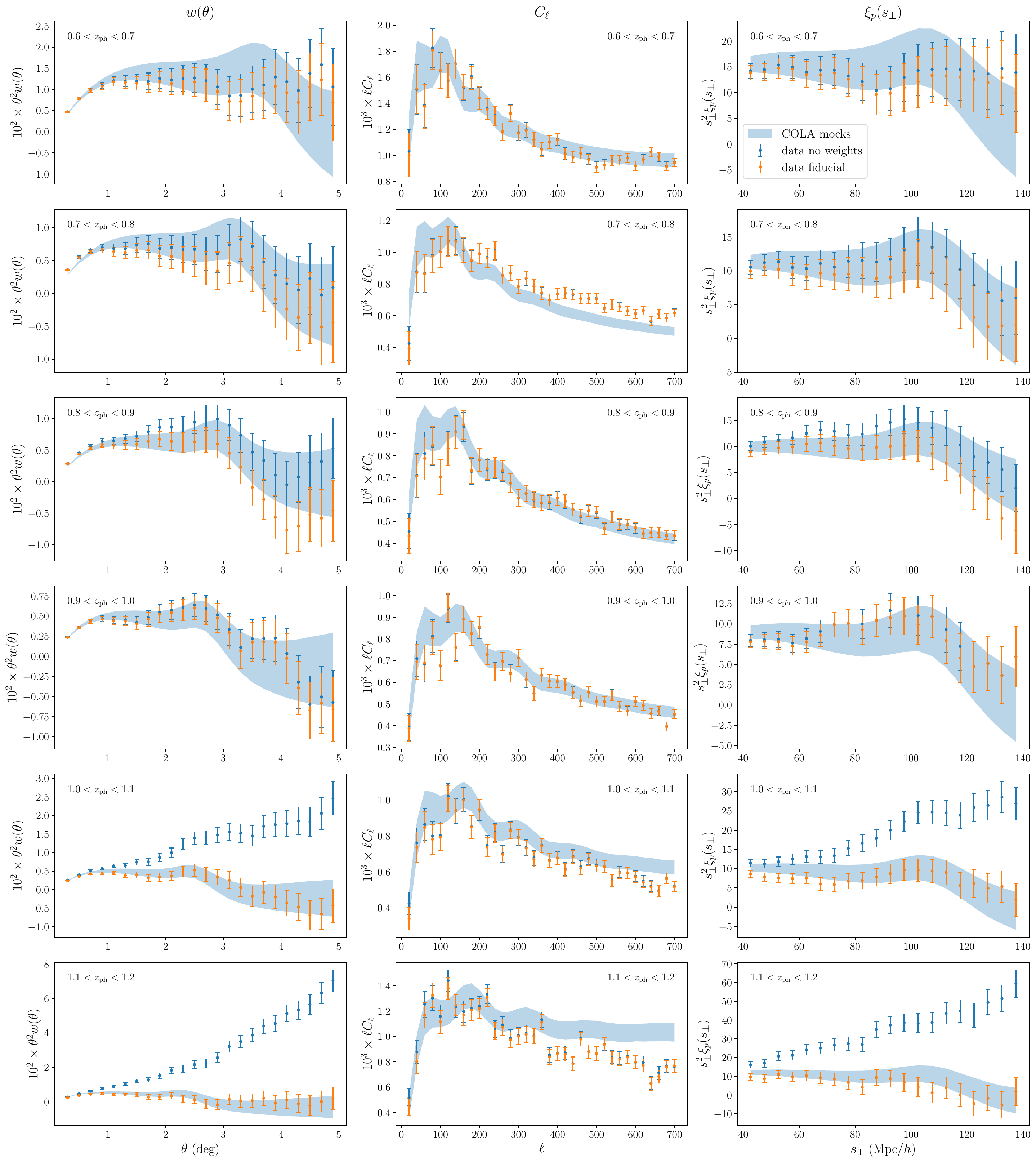}
    \caption{Clustering measurements for the 6 tomographic redshift bins of our BAO sample. The column on the left shows the angular correlation function ($w(\theta)$), the middle one shows the angular power spectra ($C_\ell$) and the one on the right shows the projected correlation function ($\xi_p(s_\perp)$). We show the results for the cases of not correcting and correcting the systematics (blue and orange points with error-bars, respectively).
    The errors were computed from the diagonals of the fiducial \cosmolike covariance matrix (or Gaussian covariance for $\xi_p$) used to run the BAO fits in \cite{Y6_BAO_measurement}.
    The blue band corresponds to the 1$\sigma$ region coming from the 1952 COLA simulations generated for our analysis (i.e., the average clustering signal of the COLA mocks $\pm$ the square root of the diagonal of their covariance), see \cite{Y6_BAO_measurement} for further details.
    }
    \label{fig:clustering_measurements_DESY6BAO}
\end{figure*}

In \autoref{fig:clustering_measurements_DESY6BAO} we show the angular correlation functions ($w(\theta)$), the angular power spectra ($C_\ell$) and the projected correlation function ($\xi_p(s_\perp)$) for the different redshift bins. These are the three different estimators that we use to measure the BAO feature in \cite{Y6_BAO_measurement}.
We include the cases of not correcting and correcting for the observational systematics (blue and orange points with error-bars, respectively), as described in \autoref{sec:weights}. The errors were computed from the diagonals of the fiducial \cosmolike covariance matrix (or Gaussian covariance for $\xi_p$) used to run the BAO fits in \cite{Y6_BAO_measurement}. The blue band corresponds to the 1$\sigma$ region computed from the 1952 COLA simulations generated for our analysis, i.e., the average clustering signal of the COLA mocks $\pm$ the square root of the diagonal of their covariance, see \cite{Y6_BAO_measurement} for further details on the simulations. The effect of the observational systematics is an increase in the amplitude of the clustering. This increase in the amplitude becomes more important with redshift, being the last two redshift bins the ones with the highest contamination.


\section{Conclusions}\label{sec:conclusions}

In this paper, we have presented the data used in the DES Y6 analysis for cosmological constraints from the measurement of the BAO distance scale. The sample selection has been optimized with respect to the one used in the Y1 and Y3 analyses: the optimal flux selection is $17.5<i<19.64+2.894z_{\rm ph}$, being $17.5<i<19+3z_{\rm ph}$ the one used for the Y1/Y3. Compared to the Y3 sample, the Y6 optimal sample has one more redshift bin, $1.1<z_{\rm ph}<1.2$, increasing its effective redshift from 0.835 to 0.867. The sample covers 4,273.42 deg${}^2$ to a depth of $i<22.5$, a very similar area to that of the Y3, which was 4,108.57 deg${}^2$. It contains 15,937,556 galaxies, compared to the 7 million galaxies of the Y3 sample. This large increase in the number of galaxies with respect to the Y3 is due to the optimized $i$-magnitude cut.
We forecast an increase in the precision of the measurement of the BAO scale of around $25\%$ with respect to the Y3 result using our Y6 optimal sample.

We have calibrated the photometric redshift distributions using VIPERS, a spectroscopic sample that has an overlapping area of 16.3 deg${}^2$ with the DES footprint and which is complete within our optimal sample selection. This allowed us to use the distributions of VIPERS \zspec as the true redshift distributions of our sample. We also used WZ estimations of the redshift distributions to compare to those of \zspec and found a very good agreement between them. Because of the noisy nature of both \zspec and WZ and because of DNF giving wider redshift distributions compared to these two, the redshift distributions used for the fiducial analysis on the data are a combination of DNF results, \zspec and WZ: DNF PDF shifted and stretched to match the properties of WZ in the first 4 bins and those of VIPERS \zspec in the last 2.

Systematics have been mitigated using the ISD algorithm, which was also the fiducial in the Y3 analysis \cite{monroy}. The residual stellar contamination, which contributes additively to the number density of our galaxy sample, has been corrected with a novel technique, which is further described in \cite{Y6_LSS_sys}. We have found a fraction of stellar contamination from 0.7\% to 3.3\% in the final BAO sample, depending on the redshift bin. This is corrected at the level of the clustering measurement, either $w(\theta)$, $C_\ell$ or $\xi_p(s_\perp)$, multiplying by $(1-f_{\rm star})^{-2}$. As a final result, we included the unblinded clustering measurements of our data ($w(\theta)$, $C_\ell$ and $\xi_p(s_\perp)$), correcting and not correcting for the observational systematics. The measurement of the BAO scale using these clustering measurements and its cosmological implications are described in \cite{Y6_BAO_measurement}.

\begin{acknowledgements}
    \input{acknowledgements}
\end{acknowledgements}

\appendix

\section{Redshift-Related Quantities}\label{app:photoz}

It is worth defining several photo-$z$-dependent quantities that we use throughout this paper:
\begin{itemize}
    \item $\langle z\rangle$. We define it as the mean photometric redshift of a given redshift bin weighted with the redshift distribution, $n(z)$, of that same redshift bin,
    \begin{equation}\label{eq:z_average}
        \langle z\rangle=\int dz\hspace{0.05cm}zn(z).
    \end{equation}
    \item $W_{68}$. We define it as the width in redshift that encloses 68\% of the integral of the redshift distribution, i.e., it is given by
    \begin{equation}\label{eq:w_68}
        W_{68}=\frac{b-a}{2}
    \end{equation}
    such that
    \begin{equation}
        \int_0^adz\hspace{0.05cm}n(z)=\int_b^\infty dz\hspace{0.05cm}n(z)=0.1585.
    \end{equation}
    \item $\sigma_{68}$ ($=\Sigma_z$). We define it as the size of the region that encloses 68\% of the distribution of 
    \begin{equation}\label{eq:sigma_68}
        \frac{\zmean-\zmc}{1+\zmc}.
    \end{equation}
    Unlike the previous two, $\sigma_{68}$ is a DNF-related quantity and cannot be computed for all our alternative estimations of the redshift distributions.
\end{itemize}

\section{Calibration of Photometric Redshifts Using VIPERS}\label{app:redshift_calibration}

Our goal here is to explicitly calibrate our photo-$z$ using VIPERS. In order to do so, we compare the redshift distributions of the BAO sample, computed with \zmc, with those of VIPERS, also computed with \zmc (we actually use the sub-sample of VIPERS matched to the BAO sample). If these two are statistically compatible, we can use the distributions of \zspec as our true redshift distributions, since VIPERS is representative of our full sample. Given that VIPERS is complete and is defined within the selection cuts of our samples, this holds true, but here we explicitly demonstrate it.

The first step to validate the photo-$z$ is to select those galaxies from VIPERS that also belong to the BAO sample. Hereafter, we refer to this sample as VIPERS for simplicity.
After matching with the BAO sample, we end up with 11,202 VIPERS galaxies, i.e., VIPERS represents, approximately, a 0.066\% of the total number of galaxies in the BAO sample. It is also worth mentioning that, in order to take into account the spectroscopic success ratio of VIPERS, which is encoded in the variable {\sc ssr} of the VIPERS catalog \cite{scodeggio2018vimos}, we must weight each VIPERS galaxy with $1/\textsc{ssr}$. 

To quantitatively compare the distributions of \zmc of the BAO sample and VIPERS (both of them shown in \autoref{fig:nz_lots} as blue histograms and blue points with error-bars, respectively), we calculate the $\chi^2$ between them 
as
\begin{equation}\label{eq:chi2_BAO_VIPERS}
    \chi^2=\sum_i\frac{\left[n_{\rm BAO}(z_i)-n_{\rm VIPERS}(z_i)\right]^2}{\Delta n_{\rm VIPERS}(z_i)^2},
\end{equation}
where the sum over $i$ means summing over histogram bins and $\Delta n(z)$ is the shot-noise contribution to the error in the redshift distributions (which is negligible for the BAO sample, because of the large number of galaxies compared to that of VIPERS).
In \autoref{tab:nz_des_vipers_chi2} we show the reduced $\chi^2$ obtained using \autoref{eq:chi2_BAO_VIPERS} and also their corresponding p-values, for each redshift bin.
All the $\chi^2$ and p-values show that both samples are compatible.
\begin{table}
    \centering
    \begin{tabular}{ccc}
        \toprule\toprule
        Bin & $\chi^2$/dof & p-value \\\hline
        $0.6<z_{\rm ph}<0.7$ & 1.29 & 0.23 \\
        $0.7<z_{\rm ph}<0.8$ & 0.87 & 0.57 \\
        $0.8<z_{\rm ph}<0.9$ & 1.08 & 0.37 \\
        $0.9<z_{\rm ph}<1.0$ & 1.68 & 0.08 \\
        $1.0<z_{\rm ph}<1.1$ & 1.54 & 0.12 \\
        $1.1<z_{\rm ph}<1.2$ & 1.07 & 0.38 \\
        \bottomrule\bottomrule
    \end{tabular}
    \caption{Reduced $\chi^2$ between the redshift distributions of \zmc of the Y6 BAO sample and VIPERS.}
    \label{tab:nz_des_vipers_chi2}
\end{table}


\section{Calibration of the Redshift Distributions for the PCF}\label{app:calibration_pcf}

In the projected correlation function (PCF) modeling, we need to utilize fine $\Delta z_{\rm ph}$ bins in order to accurately sample the true redshift distributions \cite{PCF_method_Y3}. Compared to the Y3 \cite{PCF_Y3_BAO}, in the Y6 analysis we go to higher redshifts and that makes this task even more challenging. Rather than using uniform $\Delta z_{\rm ph}$ bins as in the Y3 analysis, we adopt 22 logarithmic bins in the redshift range from 0.6 to 1.2. We calibrate these distributions following the same method as the fiducial 6-bin case discussed in \autoref{sec:shift_stretch_algorithm}, i.e., we start with the smooth DNF PDF redshift distributions and then apply our shift and stretch algorithm to calibrate them using the proxy distribution derived from the WZ technique or the matched VIPERS \zspec one. For the first 17 bins (up to $z_{\rm ph}=1.02$), they are calibrated using the WZ method, whereas for the remaining 5 bins they are calibrated using VIPERS \zspec. This is consistent with what we did in \autoref{sec:photoz}, in which we used WZ for the first 4 redshift bins (i.e., below redshift 1.0) and VIPERS \zspec for the last 2 (i.e., for redshifts between 1.0 and 1.2). 

\begin{figure}
    \centering
    \includegraphics[width=1\linewidth]{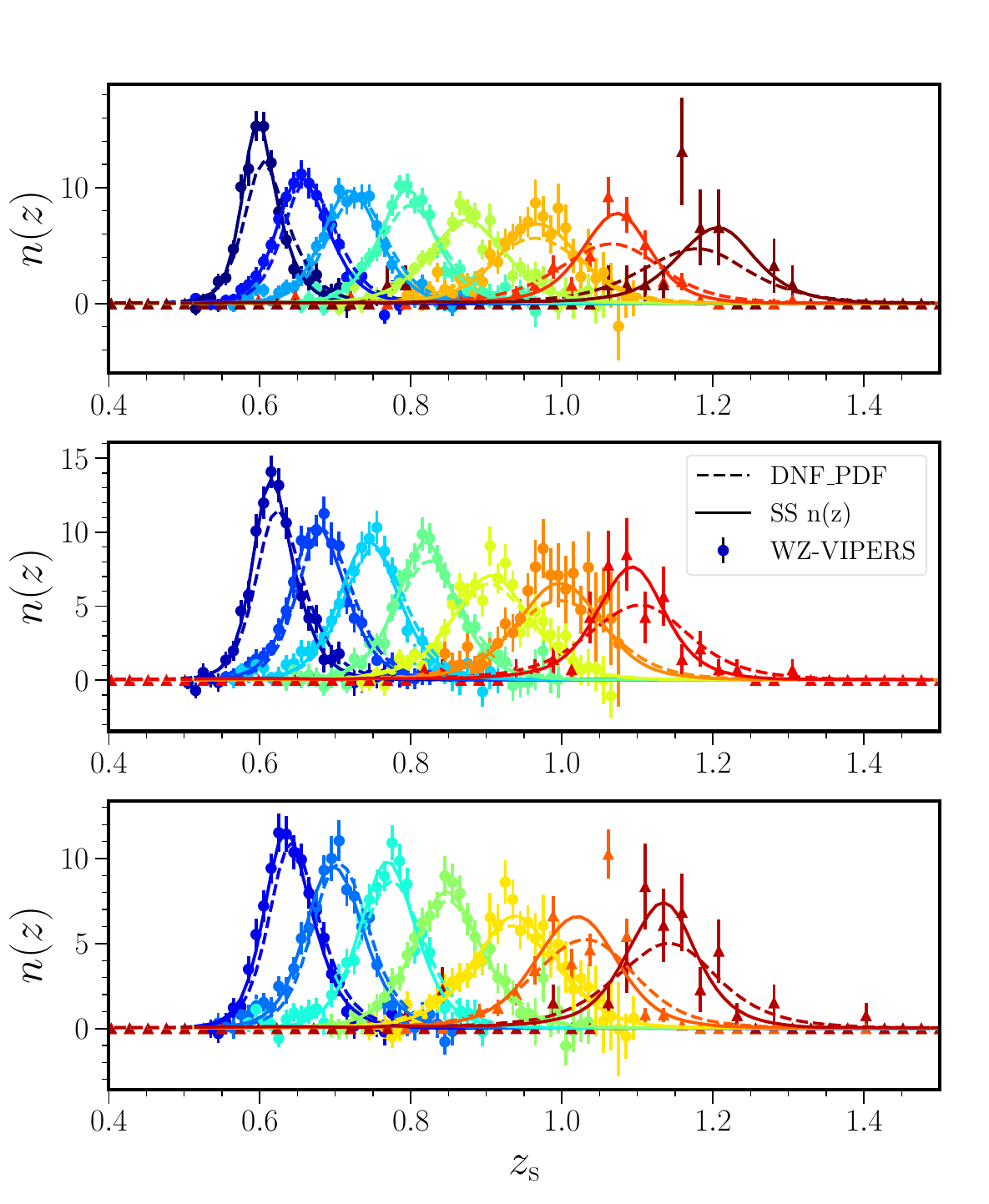}
    \caption{Redshift distributions for the 22 PCF logarithmic bins. The distributions in the upper, middle and bottom panels correspond to the bin whose bin number modulo 3 is 1, 2 and 0 respectively. The original DNF PDF (dashed) and the shifted and stretched (solid) distributions are compared. The proxy distributions used for the correction are shown in markers: circles for WZ (the first 17 bins) and triangles for matched VIPERS \zspec (the remaining 5 bins).}
    \label{fig:nz_PCFfiducial_22bins}
\end{figure}

In \autoref{fig:nz_PCFfiducial_22bins} we compare the original DNF PDF and the shifted and stretched ones. For clarity, we split the bins into three panels, with the upper, middle and bottom panels corresponding to the bin number modulo 3 being 1, 2 and 0, respectively. We have also plotted the proxy WZ or VIPERS \zspec distributions. The PDF distribution is generally wider than the calibrated one, as expected. The correction also tends to shift the distribution to a slightly lower redshift and this shift is observed in most of the redshift bins.

\section{Evolution of the 1D Relations with the Weighting Process }\label{app:1d_plots}
\begin{figure*}
    
    \begin{minipage}[b]{\textwidth}
        \centering
        \includegraphics[width=\linewidth]{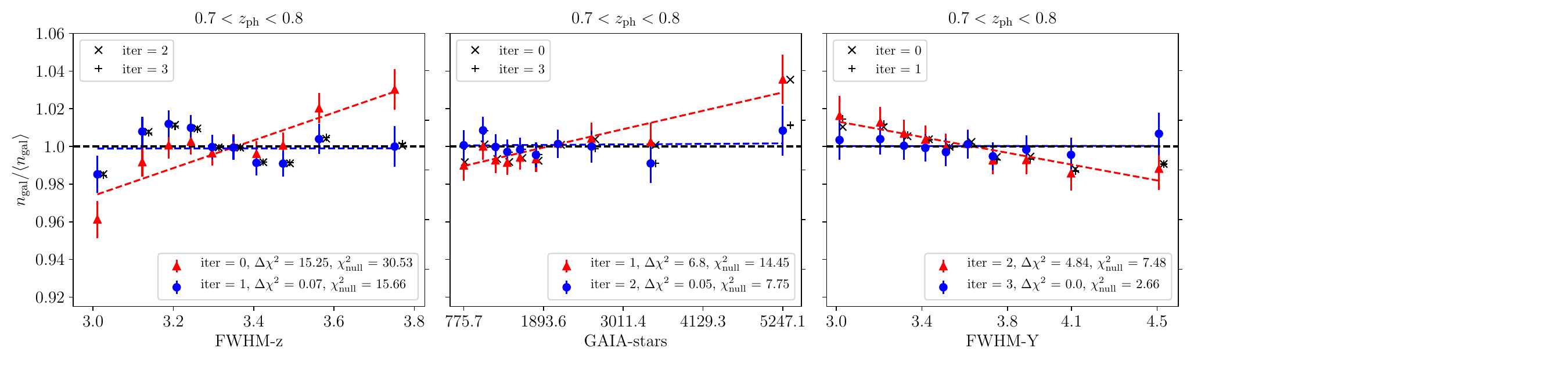}
    \end{minipage}\hfill

    \begin{minipage}[b]{\textwidth}
        \centering
        \includegraphics[width=\linewidth]{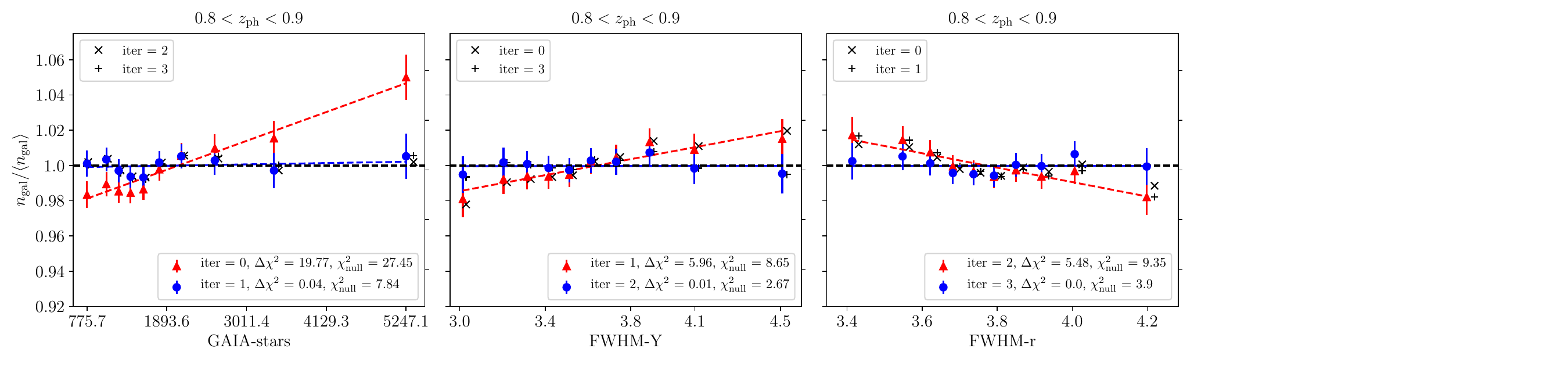}
    \end{minipage}\hfill

    \begin{minipage}[b]{\textwidth}
        \centering
        \includegraphics[width=\linewidth]{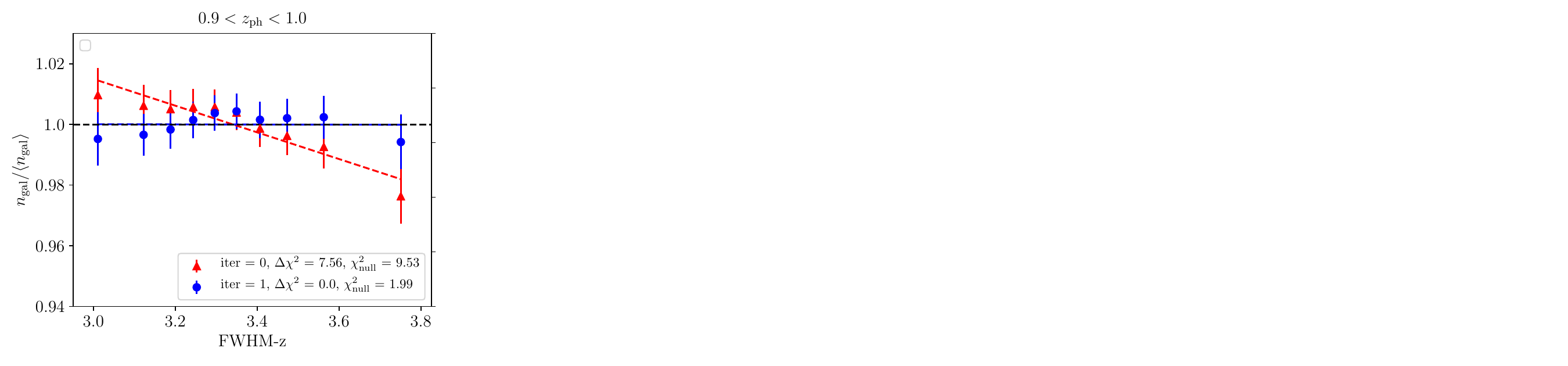}
    \end{minipage}\hfill

    \begin{minipage}[b]{\textwidth}
        \centering
        \includegraphics[width=\linewidth]{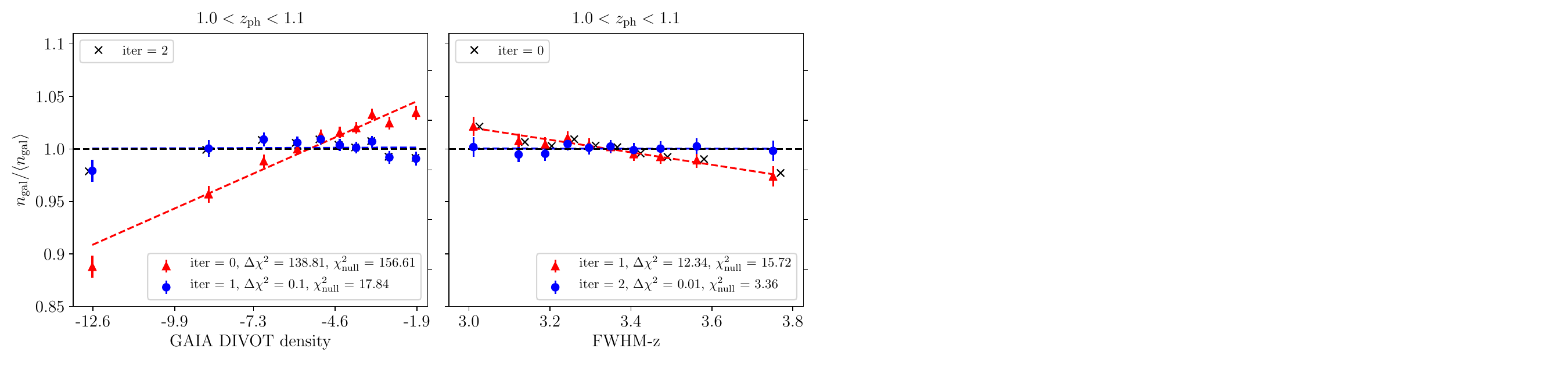}
    \end{minipage}\hfill

    \begin{minipage}[b]{\textwidth}
        \centering
        \includegraphics[width=\linewidth]{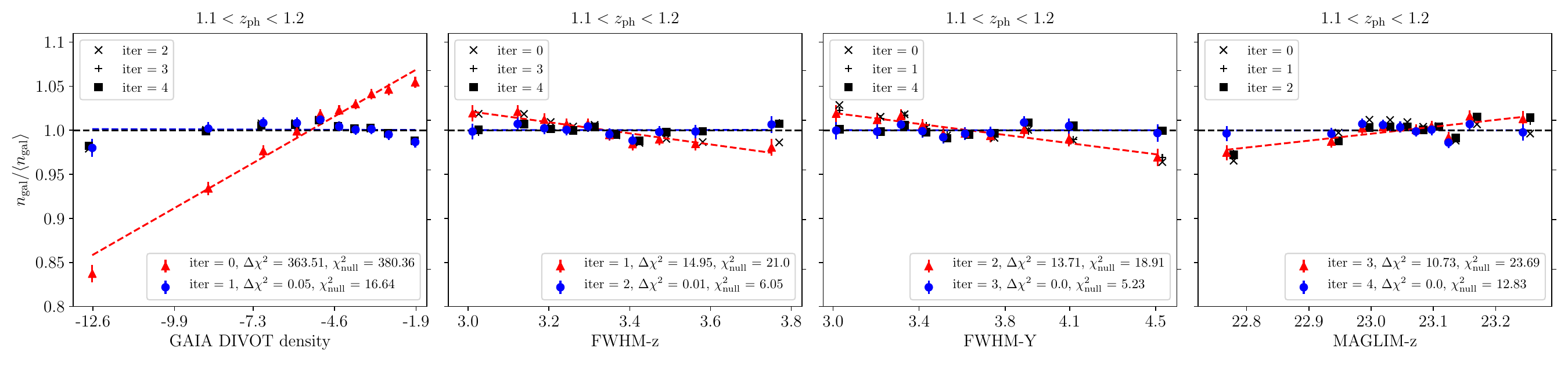}
    \end{minipage}\hfill

    \caption{Evolution of the 1D relations as a function of the correcting iteration for each SP map we correct for. Each row corresponds to a redshift bin, starting from the second one (see \autoref{fig:1d_plots} for the first bin). Red triangles show the 1D relation of each of the SP maps from \autoref{tab:spmaps_baosample} before correcting for them, blue dots correspond to their 1D relations right after correcting for them and the markers in black show their status at the intermediate iterations. Dashed lines represent the best linear fits in each case (from which $\Delta \chi^2_{\rm model}$ is computed). }
    \label{fig:all_1ds}
\end{figure*}

In this appendix, we present the evolution of the 1D relation (and, therefore, of the contamination significance) for each of the SP maps we found necessary to correct for, according to ISD. This evolution is illustrated in \autoref{fig:all_1ds}, where for each of the SP maps from \autoref{tab:spmaps_baosample} we show their 1D relation before (in red) and after (in blue) correcting for them at their corresponding iteration. We also show their status at the intermediate iterations (in black). 

\section{Weights Validation Tests}\label{app:wval_tests}

\begin{figure*}
    \begin{minipage}[b]{0.49\textwidth}
        \centering
        \includegraphics[width=\linewidth]{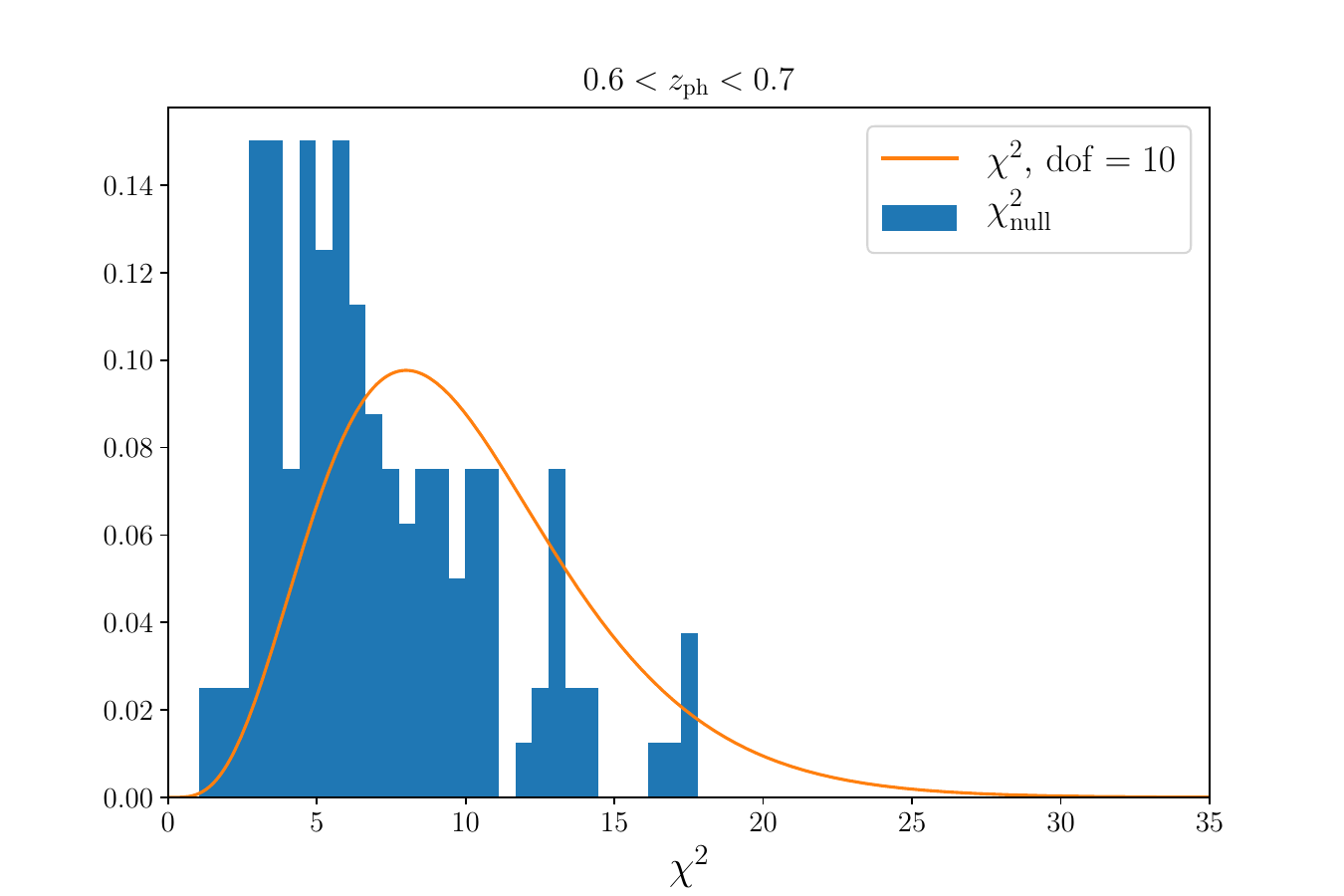}
    \end{minipage}\hfill
    \begin{minipage}[b]{0.49\textwidth}
        \centering
        \includegraphics[width=\linewidth]{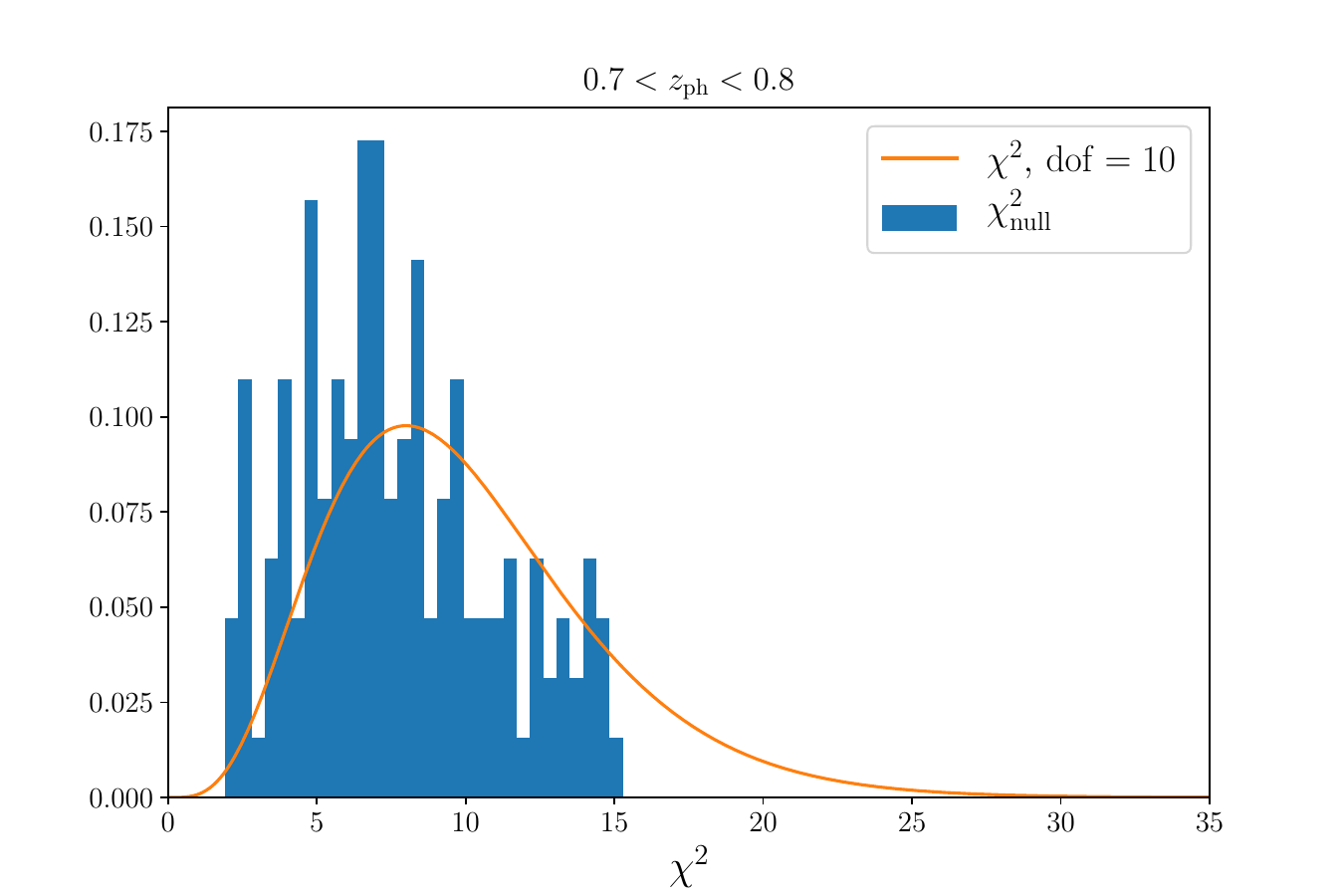}
    \end{minipage}\hfill
    \begin{minipage}[b]{0.49\textwidth}
        \centering
        \includegraphics[width=\linewidth]{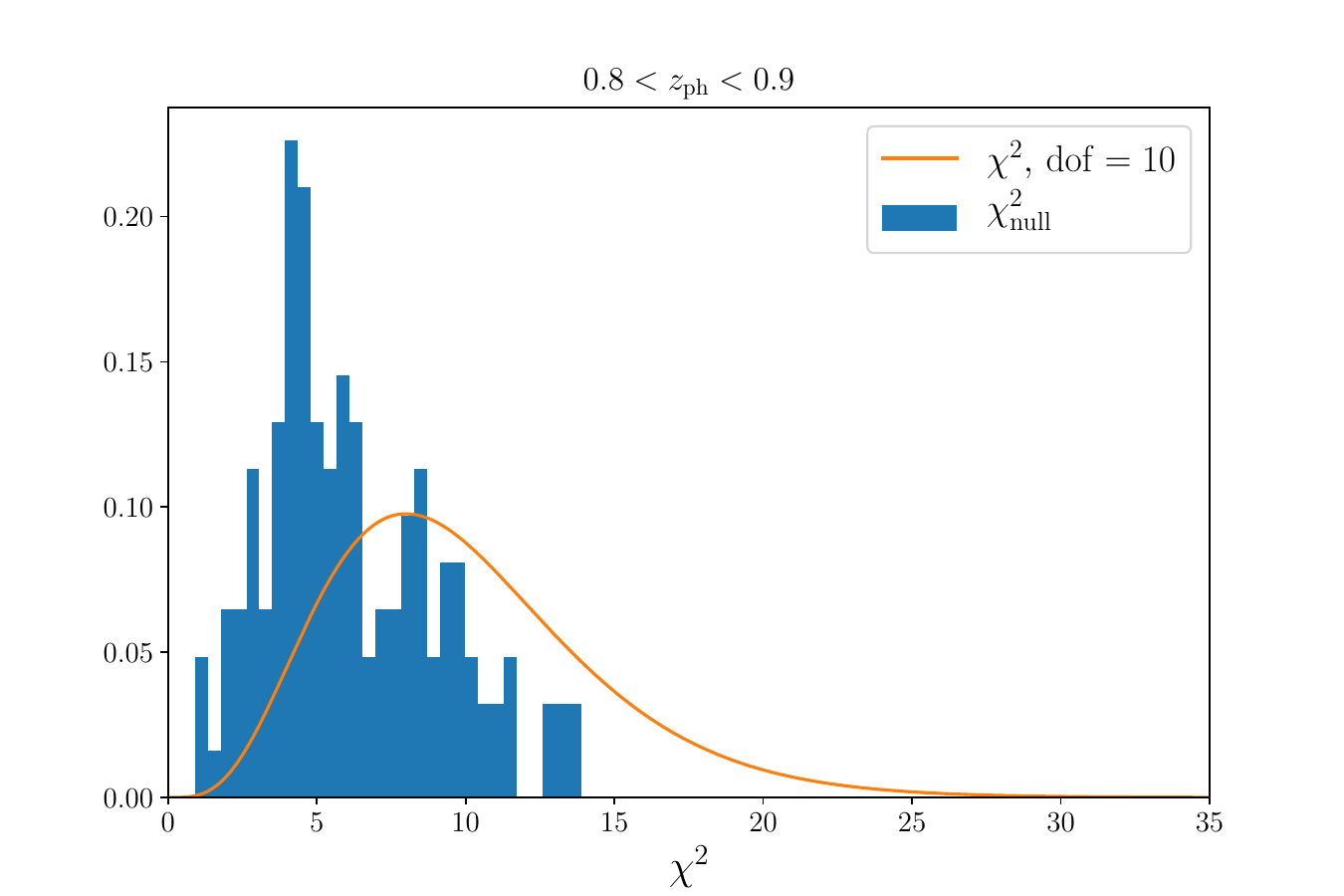}
    \end{minipage}
    \begin{minipage}[b]{0.49\textwidth}
        \centering
        \includegraphics[width=\linewidth]{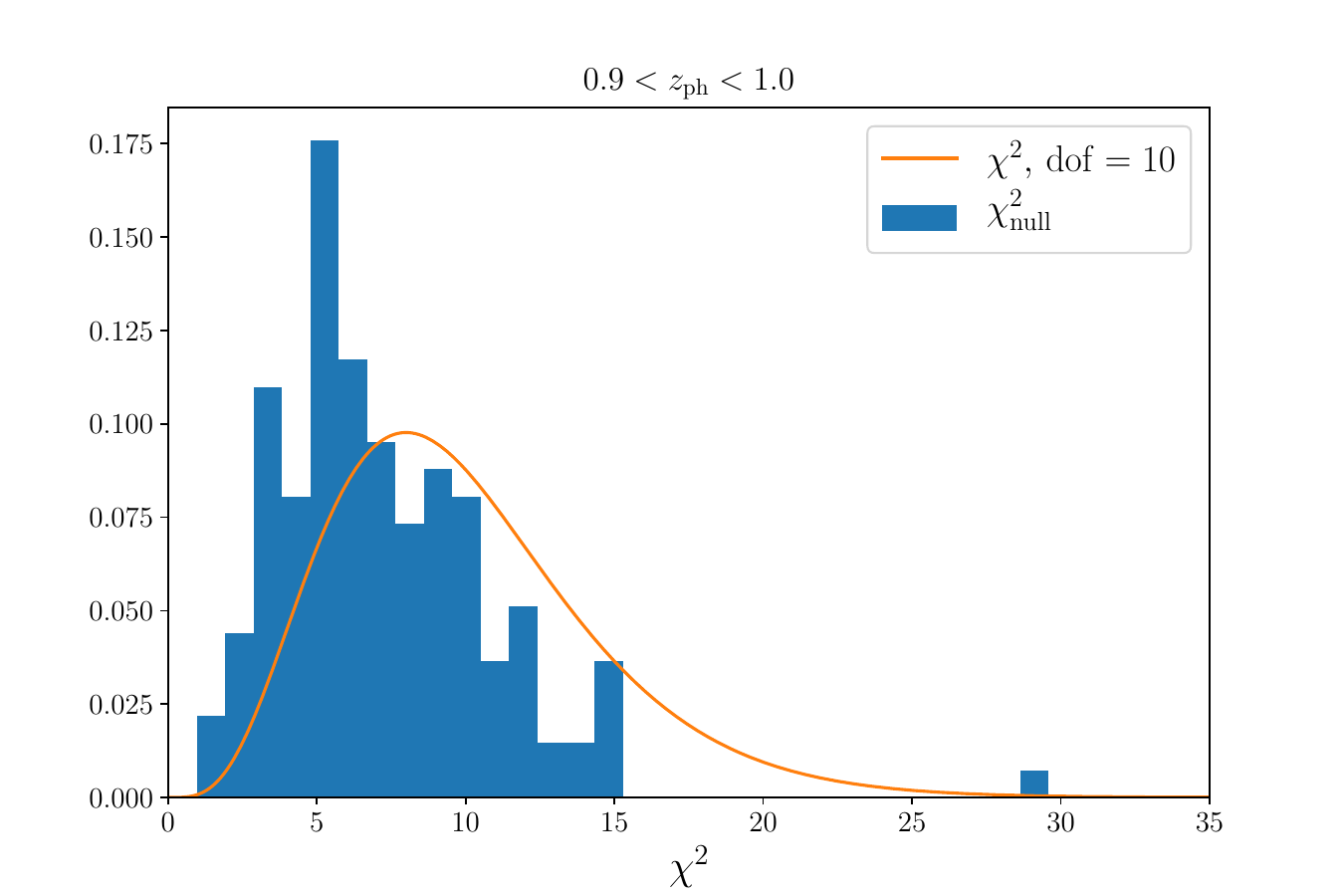}
    \end{minipage}\hfill
    \begin{minipage}[b]{0.49\textwidth}
        \centering
        \includegraphics[width=\linewidth]{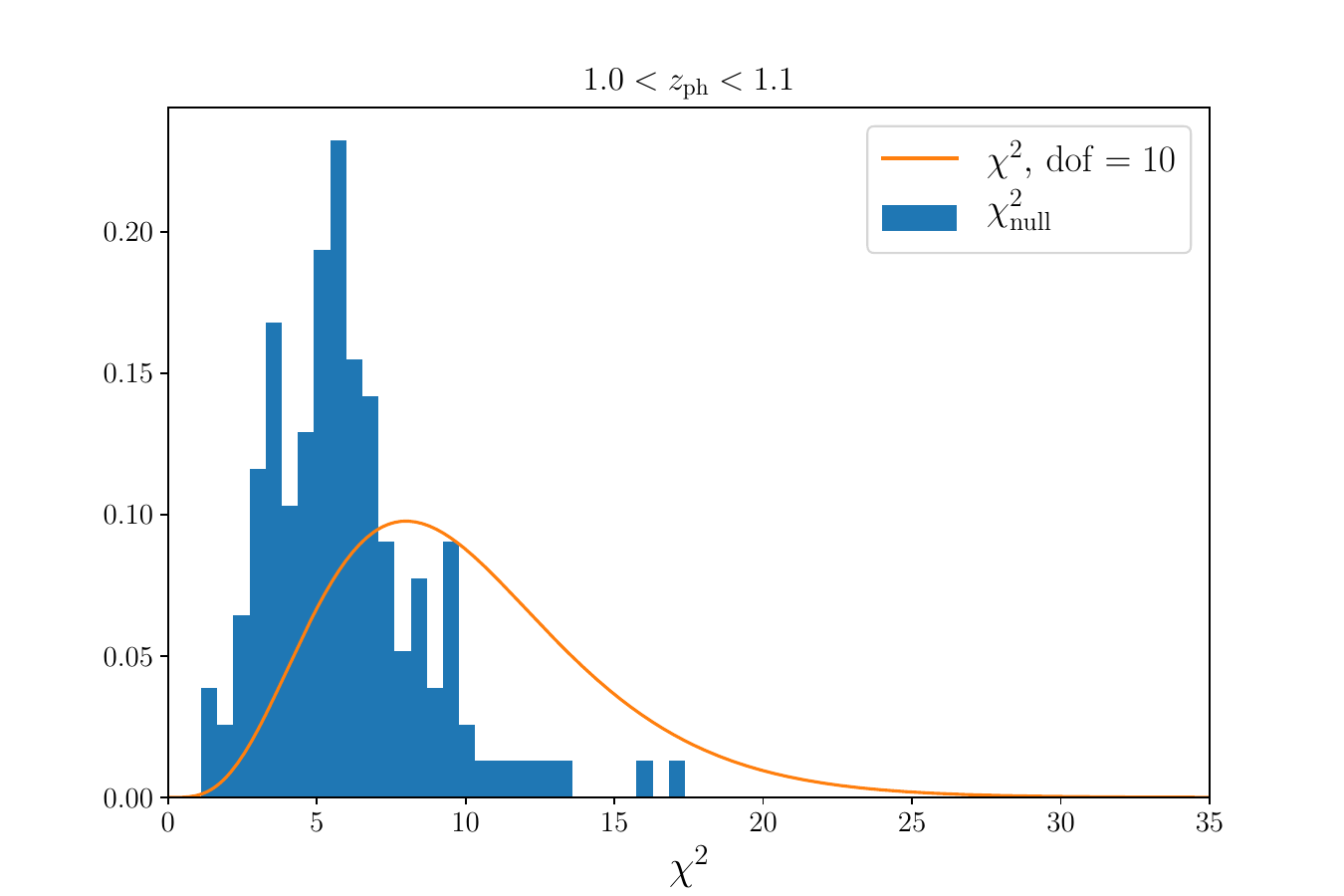}
    \end{minipage}\hfill
    \begin{minipage}[b]{0.49\textwidth}
        \centering
        \includegraphics[width=\linewidth]{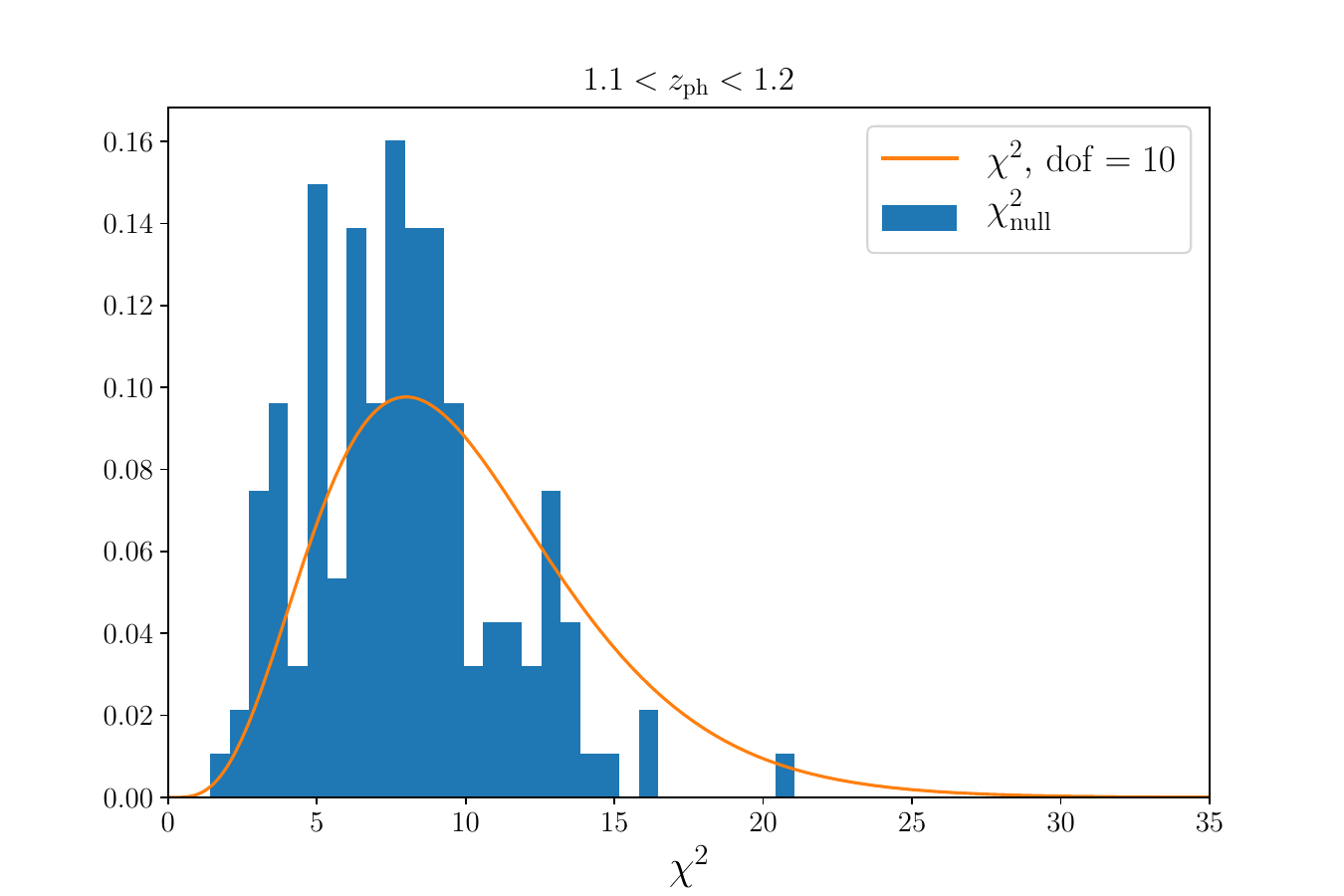}
    \end{minipage}
    \caption{Distribution of $\chi^2_{\rm null}$ computed on the weighted Y6 BAO sample for all available SP maps. The solid orange line showcases a theoretical $\chi^2$ distribution with $N_{\rm dof} = 10$. We find no significant deviations of $\chi^2_{\rm null}$ from the theoretical distribution, meaning that the linear assumption for the systematic correction is valid to the extent, i.e., significance threshold, that we consider. }
    \label{fig:chi2_null}
\end{figure*}

To check the correct functioning of ISD and the weight obtained with it, we run this method again on the weighted BAO sample using the fiducial set of SP maps presented in \autoref{sec:weights}. In doing so, we simply evaluate the correct functioning of the method, since by definition all those SP maps should be found to have $S_{\rm 1D} < 2$, so no additional corrections should be needed. We find no remaining levels of contamination at $S_{\rm 1D} > 2$ coming from any of the fiducial SP maps. After this, we run ISD on the weighted BAO sample, this time using the full list of available SP maps in Y6 (see \cite{Y6_gold}) and \cite{Y6_LSS_sys}), that is, without limiting the list of contamination templates to the fiducial set presented in \autoref{sec:weights}. Proceeding this way, we test the validity of our fiducial set of SP maps as a representative set of contamination templates. With this configuration, ISD finds minor levels of additional contamination at some redshift bins in the form of additional (1 to 3) SP maps to be corrected for. However, these numbers are compatible with statistical fluctuations around a strict significance threshold and, given the negligible impact of the weights on the BAO peak, we decided not to incorporate those maps in the final set of corrections. Lastly, we test the assumption of linearity for the corrections by comparing the distribution of $\chi^2_{\rm null}$ measured on the weighted BAO sample for all SP maps with a theoretical $\chi^2$ distribution with $N_{\rm dof} = 10$ (the number of bins used for the 1D relations). Deviations from linearity of the 1D relations and, therefore, of the corresponding (linear) corrections should appear as deviations from a $\chi^2$ behavior of $\chi^2_{\rm null}$. The results of this test are depicted in \autoref{fig:chi2_null}. We conclude that the systematic corrections provided by ISD show no significant deviations from linearity. At the fourth and sixth redshift bins, we observe two outlier values of $\chi^2_{\rm null}$, but checking the 1D relations of those two SP maps we find them still compatible with linearity ($\chi^2_{\rm model} = 11.6$ and $14.1$ with $N_{\rm dof} = 8$). Their slightly outlier values are explained by mild levels of residual contamination not higher than $\sim 4\%$ at the 1D level, which ISD does not flag as significant enough. 

\bibliography{y6kp_baosample_des}

\end{document}

%% file: authorlist.tex
\author{J. Mena-Fern\'andez}\email{juan.menafernandez@lpsc.in2p3.fr}
\affiliation{Centro de Investigaciones Energ\'eticas, Medioambientales y Tecnol\'ogicas (CIEMAT), Madrid, Spain}
\affiliation{Universit\'e Grenoble Alpes, CNRS, LPSC-IN2P3, 38000 Grenoble, France}
\author{M.~Rodr\'iguez-Monroy}
\affiliation{Laboratoire de physique des 2 infinis Ir\`ene Joliot-Curie, CNRS Universit\'e Paris-Saclay, Bât. 100, F-91405 Orsay Cedex, France}
\affiliation{Instituto de Fisica Teorica UAM/CSIC, Universidad Autonoma de Madrid, 28049 Madrid, Spain}
\author{S.~Avila}
\affiliation{Institut de F\'{\i}sica d'Altes Energies (IFAE), The Barcelona Institute of Science and Technology, Campus UAB, 08193 Bellaterra (Barcelona) Spain}
\author{A.~Porredon}
\affiliation{Ruhr University Bochum, Faculty of Physics and Astronomy, Astronomical Institute, German Centre for Cosmological Lensing, 44780 Bochum, Germany}
\author{K.~C.~Chan}
\affiliation{School of Physics and Astronomy, Sun Yat-sen University, 2 Daxue Road, Tangjia, Zhuhai 519082, China}
\affiliation{CSST Science Center for the Guangdong-Hongkong-Macau Greater Bay Area, SYSU, Zhuhai 519082, China}
\author{H.~Camacho}
\affiliation{Instituto de F\'{i}sica Te\'orica, Universidade Estadual Paulista, S\~ao Paulo, Brazil}
\affiliation{Laborat\'orio Interinstitucional de e-Astronomia - LIneA, Rua Gal. Jos\'e Cristino 77, Rio de Janeiro, RJ - 20921-400, Brazil}
\affiliation{Brookhaven National Laboratory, Bldg 510, Upton, NY 11973, USA}
\author{N.~Weaverdyck}
\affiliation{Department of Physics, University of Michigan, Ann Arbor, MI 48109, USA}
\affiliation{Lawrence Berkeley National Laboratory, 1 Cyclotron Road, Berkeley, CA 94720, USA}
\author{I.~Sevilla-Noarbe}
\affiliation{Centro de Investigaciones Energ\'eticas, Medioambientales y Tecnol\'ogicas (CIEMAT), Madrid, Spain}
\author{E.~Sanchez}
\affiliation{Centro de Investigaciones Energ\'eticas, Medioambientales y Tecnol\'ogicas (CIEMAT), Madrid, Spain}
\author{L.~Toribio San Cipriano}
\affiliation{Centro de Investigaciones Energ\'eticas, Medioambientales y Tecnol\'ogicas (CIEMAT), Madrid, Spain}
\author{J.~De~Vicente}
\affiliation{Centro de Investigaciones Energ\'eticas, Medioambientales y Tecnol\'ogicas (CIEMAT), Madrid, Spain}
\author{I.~Ferrero}
\affiliation{Institute of Theoretical Astrophysics, University of Oslo. P.O. Box 1029 Blindern, NO-0315 Oslo, Norway}
\author{R.~Cawthon}
\affiliation{Physics Department, William Jewell College, Liberty, MO, 64068}
\author{A.~Carnero~Rosell}
\affiliation{Instituto de Astrofisica de Canarias, E-38205 La Laguna, Tenerife, Spain}
\affiliation{Laborat\'orio Interinstitucional de e-Astronomia - LIneA, Rua Gal. Jos\'e Cristino 77, Rio de Janeiro, RJ - 20921-400, Brazil}
\affiliation{Universidad de La Laguna, Dpto. Astrofísica, E-38206 La Laguna, Tenerife, Spain}
\author{J.~Elvin-Poole}
\affiliation{Department of Physics and Astronomy, University of Waterloo, 200 University Ave W, Waterloo, ON N2L 3G1, Canada}
\author{G.~Giannini}
\affiliation{Institut de F\'{\i}sica d'Altes Energies (IFAE), The Barcelona Institute of Science and Technology, Campus UAB, 08193 Bellaterra (Barcelona) Spain}
\affiliation{Kavli Institute for Cosmological Physics, University of Chicago, Chicago, IL 60637, USA}
\author{S.~Lee}
\affiliation{Jet Propulsion Laboratory, California Institute of Technology, 4800 Oak Grove Dr., Pasadena, CA 91109, USA}
\author{M.~Adamow}
\affiliation{Center for Astrophysical Surveys, National Center for Supercomputing Applications, 1205 West Clark St., Urbana, IL 61801, USA}
\author{K.~Bechtol}
\affiliation{Physics Department, 2320 Chamberlin Hall, University of Wisconsin-Madison, 1150 University Avenue Madison, WI  53706-1390}
\author{A.~Drlica-Wagner}
\affiliation{Department of Astronomy and Astrophysics, University of Chicago, Chicago, IL 60637, USA}
\affiliation{Fermi National Accelerator Laboratory, P. O. Box 500, Batavia, IL 60510, USA}
\affiliation{Kavli Institute for Cosmological Physics, University of Chicago, Chicago, IL 60637, USA}
\author{R.~A.~Gruendl}
\affiliation{Center for Astrophysical Surveys, National Center for Supercomputing Applications, 1205 West Clark St., Urbana, IL 61801, USA}
\affiliation{Department of Astronomy, University of Illinois at Urbana-Champaign, 1002 W. Green Street, Urbana, IL 61801, USA}
\author{W.~G.~Hartley}
\affiliation{Department of Astronomy, University of Geneva, ch. d'\'Ecogia 16, CH-1290 Versoix, Switzerland}
\author{A.~Pieres}
\affiliation{Laborat\'orio Interinstitucional de e-Astronomia - LIneA, Rua Gal. Jos\'e Cristino 77, Rio de Janeiro, RJ - 20921-400, Brazil}
\affiliation{Observat\'orio Nacional, Rua Gal. Jos\'e Cristino 77, Rio de Janeiro, RJ - 20921-400, Brazil}
\author{A.~J.~Ross}
\affiliation{Center for Cosmology and Astro-Particle Physics, The Ohio State University, Columbus, OH 43210, USA}
\author{E.~S.~Rykoff}
\affiliation{Kavli Institute for Particle Astrophysics \& Cosmology, P. O. Box 2450, Stanford University, Stanford, CA 94305, USA}
\affiliation{SLAC National Accelerator Laboratory, Menlo Park, CA 94025, USA}
\author{E.~Sheldon}
\affiliation{Brookhaven National Laboratory, Bldg 510, Upton, NY 11973, USA}
\author{B.~Yanny}
\affiliation{Fermi National Accelerator Laboratory, P. O. Box 500, Batavia, IL 60510, USA}

\author{T.~M.~C.~Abbott}
\affiliation{Cerro Tololo Inter-American Observatory, NSF's National Optical-Infrared Astronomy Research Laboratory, Casilla 603, La Serena, Chile}
\author{M.~Aguena}
\affiliation{Laborat\'orio Interinstitucional de e-Astronomia - LIneA, Rua Gal. Jos\'e Cristino 77, Rio de Janeiro, RJ - 20921-400, Brazil}
\author{S.~Allam}
\affiliation{Fermi National Accelerator Laboratory, P. O. Box 500, Batavia, IL 60510, USA}
\author{O.~Alves}
\affiliation{Department of Physics, University of Michigan, Ann Arbor, MI 48109, USA}
\author{A.~Amon}
\affiliation{Institute of Astronomy, University of Cambridge, Madingley Road, Cambridge CB3 0HA, UK}
\affiliation{Kavli Institute for Cosmology, University of Cambridge, Madingley Road, Cambridge CB3 0HA, UK}
\affiliation{Department of Astrophysical Sciences, Princeton University, Peyton Hall, Princeton, NJ 08544, USA}
\author{F.~Andrade-Oliveira}
\affiliation{Department of Physics, University of Michigan, Ann Arbor, MI 48109, USA}
\author{J.~Annis}
\affiliation{Fermi National Accelerator Laboratory, P. O. Box 500, Batavia, IL 60510, USA}
\author{D.~Bacon}
\affiliation{Institute of Cosmology and Gravitation, University of Portsmouth, Portsmouth, PO1 3FX, UK}
\author{J.~Blazek}
\affiliation{Department of Physics, Northeastern University, Boston, MA 02115, USA}
\author{S.~Bocquet}
\affiliation{University Observatory, Faculty of Physics, Ludwig-Maximilians-Universit\"at, Scheinerstr. 1, 81679 Munich, Germany}
\author{D.~Brooks}
\affiliation{Department of Physics \& Astronomy, University College London, Gower Street, London, WC1E 6BT, UK}
\author{J.~Carretero}
\affiliation{Institut de F\'{\i}sica d'Altes Energies (IFAE), The Barcelona Institute of Science and Technology, Campus UAB, 08193 Bellaterra (Barcelona) Spain}
\author{F.~J.~Castander}
\affiliation{Institut d'Estudis Espacials de Catalunya (IEEC), 08034 Barcelona, Spain}
\affiliation{Institute of Space Sciences (ICE, CSIC),  Campus UAB, Carrer de Can Magrans, s/n,  08193 Barcelona, Spain}
\author{C.~Conselice}
\affiliation{Jodrell Bank Center for Astrophysics, School of Physics and Astronomy, University of Manchester, Oxford Road, Manchester, M13 9PL, UK}
\affiliation{University of Nottingham, School of Physics and Astronomy, Nottingham NG7 2RD, UK}
\author{M.~Crocce}
\affiliation{Institut d'Estudis Espacials de Catalunya (IEEC), 08034 Barcelona, Spain}
\affiliation{Institute of Space Sciences (ICE, CSIC),  Campus UAB, Carrer de Can Magrans, s/n,  08193 Barcelona, Spain}
\author{L.~N.~da Costa}
\affiliation{Laborat\'orio Interinstitucional de e-Astronomia - LIneA, Rua Gal. Jos\'e Cristino 77, Rio de Janeiro, RJ - 20921-400, Brazil}
\author{M.~E.~S.~Pereira}
\affiliation{Hamburger Sternwarte, Universit\"{a}t Hamburg, Gojenbergsweg 112, 21029 Hamburg, Germany}
\author{T.~M.~Davis}
\affiliation{School of Mathematics and Physics, University of Queensland,  Brisbane, QLD 4072, Australia}
\author{N.~Deiosso}
\affiliation{Centro de Investigaciones Energ\'eticas, Medioambientales y Tecnol\'ogicas (CIEMAT), Madrid, Spain}
\author{S.~Desai}
\affiliation{Department of Physics, IIT Hyderabad, Kandi, Telangana 502285, India}
\author{H.~T.~Diehl}
\affiliation{Fermi National Accelerator Laboratory, P. O. Box 500, Batavia, IL 60510, USA}
\author{S.~Dodelson}
\affiliation{Department of Physics, Carnegie Mellon University, Pittsburgh, Pennsylvania 15312, USA}
\affiliation{NSF AI Planning Institute for Physics of the Future, Carnegie Mellon University, Pittsburgh, PA 15213, USA}
\author{C.~Doux}
\affiliation{Department of Physics and Astronomy, University of Pennsylvania, Philadelphia, PA 19104, USA}
\affiliation{Universit\'e Grenoble Alpes, CNRS, LPSC-IN2P3, 38000 Grenoble, France}
\author{S.~Everett}
\affiliation{Jet Propulsion Laboratory, California Institute of Technology, 4800 Oak Grove Dr., Pasadena, CA 91109, USA}
\author{J.~Frieman}
\affiliation{Fermi National Accelerator Laboratory, P. O. Box 500, Batavia, IL 60510, USA}
\affiliation{Kavli Institute for Cosmological Physics, University of Chicago, Chicago, IL 60637, USA}
\author{J.~Garc\'ia-Bellido}
\affiliation{Instituto de Fisica Teorica UAM/CSIC, Universidad Autonoma de Madrid, 28049 Madrid, Spain}
\author{E.~Gaztanaga}
\affiliation{Institut d'Estudis Espacials de Catalunya (IEEC), 08034 Barcelona, Spain}
\affiliation{Institute of Cosmology and Gravitation, University of Portsmouth, Portsmouth, PO1 3FX, UK}
\affiliation{Institute of Space Sciences (ICE, CSIC),  Campus UAB, Carrer de Can Magrans, s/n,  08193 Barcelona, Spain}
\author{G.~Gutierrez}
\affiliation{Fermi National Accelerator Laboratory, P. O. Box 500, Batavia, IL 60510, USA}
\author{S.~R.~Hinton}
\affiliation{School of Mathematics and Physics, University of Queensland,  Brisbane, QLD 4072, Australia}
\author{D.~L.~Hollowood}
\affiliation{Santa Cruz Institute for Particle Physics, Santa Cruz, CA 95064, USA}
\author{K.~Honscheid}
\affiliation{Center for Cosmology and Astro-Particle Physics, The Ohio State University, Columbus, OH 43210, USA}
\affiliation{Department of Physics, The Ohio State University, Columbus, OH 43210, USA}
\author{D.~Huterer}
\affiliation{Department of Physics, University of Michigan, Ann Arbor, MI 48109, USA}
\author{K.~Kuehn}
\affiliation{Australian Astronomical Optics, Macquarie University, North Ryde, NSW 2113, Australia}
\affiliation{Lowell Observatory, 1400 Mars Hill Rd, Flagstaff, AZ 86001, USA}
\author{O.~Lahav}
\affiliation{Department of Physics \& Astronomy, University College London, Gower Street, London, WC1E 6BT, UK}
\author{C.~Lidman}
\affiliation{Centre for Gravitational Astrophysics, College of Science, The Australian National University, ACT 2601, Australia}
\affiliation{The Research School of Astronomy and Astrophysics, Australian National University, ACT 2601, Australia}
\author{H.~Lin}
\affiliation{Fermi National Accelerator Laboratory, P. O. Box 500, Batavia, IL 60510, USA}
\author{J.~L.~Marshall}
\affiliation{George P. and Cynthia Woods Mitchell Institute for Fundamental Physics and Astronomy, and Department of Physics and Astronomy, Texas A\&M University, College Station, TX 77843,  USA}
\author{F.~Menanteau}
\affiliation{Center for Astrophysical Surveys, National Center for Supercomputing Applications, 1205 West Clark St., Urbana, IL 61801, USA}
\affiliation{Department of Astronomy, University of Illinois at Urbana-Champaign, 1002 W. Green Street, Urbana, IL 61801, USA}
\author{R.~Miquel}
\affiliation{Instituci\'o Catalana de Recerca i Estudis Avan\c{c}ats, E-08010 Barcelona, Spain}
\affiliation{Institut de F\'{\i}sica d'Altes Energies (IFAE), The Barcelona Institute of Science and Technology, Campus UAB, 08193 Bellaterra (Barcelona) Spain}
\author{J.~Myles}
\affiliation{Department of Astrophysical Sciences, Princeton University, Peyton Hall, Princeton, NJ 08544, USA}
\author{R.~L.~C.~Ogando}
\affiliation{Observat\'orio Nacional, Rua Gal. Jos\'e Cristino 77, Rio de Janeiro, RJ - 20921-400, Brazil}
\author{A.~Palmese}
\affiliation{Department of Physics, Carnegie Mellon University, Pittsburgh, Pennsylvania 15312, USA}
\author{W.~J.~Percival}
\affiliation{Department of Physics and Astronomy, University of Waterloo, 200 University Ave W, Waterloo, ON N2L 3G1, Canada}
\affiliation{Perimeter Institute for Theoretical Physics, 31 Caroline St. North, Waterloo, ON N2L 2Y5, Canada}
\author{A.~A.~Plazas~Malag\'on}
\affiliation{Kavli Institute for Particle Astrophysics \& Cosmology, P. O. Box 2450, Stanford University, Stanford, CA 94305, USA}
\affiliation{SLAC National Accelerator Laboratory, Menlo Park, CA 94025, USA}
\author{A.~Roodman}
\affiliation{Kavli Institute for Particle Astrophysics \& Cosmology, P. O. Box 2450, Stanford University, Stanford, CA 94305, USA}
\affiliation{SLAC National Accelerator Laboratory, Menlo Park, CA 94025, USA}
\author{R.~Rosenfeld}
\affiliation{ICTP South American Institute for Fundamental Research\\ Instituto de F\'{\i}sica Te\'orica, Universidade Estadual Paulista, S\~ao Paulo, Brazil}
\affiliation{Laborat\'orio Interinstitucional de e-Astronomia - LIneA, Rua Gal. Jos\'e Cristino 77, Rio de Janeiro, RJ - 20921-400, Brazil}
\author{S.~Samuroff}
\affiliation{Department of Physics, Northeastern University, Boston, MA 02115, USA}
\author{D.~Sanchez Cid}
\affiliation{Centro de Investigaciones Energ\'eticas, Medioambientales y Tecnol\'ogicas (CIEMAT), Madrid, Spain}
\author{B.~Santiago}
\affiliation{Instituto de F\'\i sica, UFRGS, Caixa Postal 15051, Porto Alegre, RS - 91501-970, Brazil}
\affiliation{Laborat\'orio Interinstitucional de e-Astronomia - LIneA, Rua Gal. Jos\'e Cristino 77, Rio de Janeiro, RJ - 20921-400, Brazil}
\author{M.~Schubnell}
\affiliation{Department of Physics, University of Michigan, Ann Arbor, MI 48109, USA}
\author{M.~Smith}
\affiliation{School of Physics and Astronomy, University of Southampton,  Southampton, SO17 1BJ, UK}
\author{E.~Suchyta}
\affiliation{Computer Science and Mathematics Division, Oak Ridge National Laboratory, Oak Ridge, TN 37831}
\author{M.~E.~C.~Swanson}
\affiliation{Center for Astrophysical Surveys, National Center for Supercomputing Applications, 1205 West Clark St., Urbana, IL 61801, USA}
\author{G.~Tarle}
\affiliation{Department of Physics, University of Michigan, Ann Arbor, MI 48109, USA}
\author{D.~Thomas}
\affiliation{Institute of Cosmology and Gravitation, University of Portsmouth, Portsmouth, PO1 3FX, UK}
\affiliation{School of Mathematics and Physics, University of Portsmouth, Lion Gate Building, Portsmouth, PO1 3HF, UK}
\author{C.~To}
\affiliation{Center for Cosmology and Astro-Particle Physics, The Ohio State University, Columbus, OH 43210, USA}
\author{D.~L.~Tucker}
\affiliation{Fermi National Accelerator Laboratory, P. O. Box 500, Batavia, IL 60510, USA}
\author{A.~R.~Walker}
\affiliation{Cerro Tololo Inter-American Observatory, NSF's National Optical-Infrared Astronomy Research Laboratory, Casilla 603, La Serena, Chile}
\author{J.~Weller}
\affiliation{Max Planck Institute for Extraterrestrial Physics, Giessenbachstrasse, 85748 Garching, Germany}
\affiliation{Universit\"ats-Sternwarte, Fakult\"at f\"ur Physik, Ludwig-Maximilians Universit\"at M\"unchen, Scheinerstr. 1, 81679 M\"unchen, Germany}
\author{P.~Wiseman}
\affiliation{School of Physics and Astronomy, University of Southampton,  Southampton, SO17 1BJ, UK}
\author{M.~Yamamoto}
\affiliation{Department of Physics, Duke University Durham, NC 27708, USA}

\collaboration{DES Collaboration}

%% file: abstract.tex
In this paper we present and validate the galaxy sample used for the analysis of the baryon acoustic oscillation (BAO) signal in the Dark Energy Survey (DES) Y6 data. The definition is based on a color and redshift-dependent magnitude cut optimized to select galaxies at redshifts higher than 0.6, while ensuring a high-quality photo-$z$ determination. The optimization is performed using a Fisher forecast algorithm, finding the optimal $i$-magnitude cut to be given by $i<19.64+2.894z_{\rm ph}$. For the optimal sample, we forecast an increase in precision in the BAO measurement of $\sim$25\% with respect to the Y3 analysis. Our BAO sample has a total of 15,937,556 galaxies in the redshift range $0.6<z_{\rm ph}<1.2$ and its angular mask covers 4,273.42 deg${}^2$ to a depth of $i=22.5$. We validate its redshift distributions with three different methods: directional neighborhood fitting algorithm (DNF), which is our primary photo-$z$ estimation; direct calibration with VIPERS, which is a spectroscopic galaxy sample that overlaps with our BAO sample and is complete within our selection cuts; and clustering redshift using SDSS galaxies. The fiducial redshift distribution is a combination of these three techniques performed by modifying the mean and width of the DNF distributions to match those of VIPERS and clustering redshift. In this paper we also describe the methodology used to mitigate the effect of observational systematics, which is analogous to the one used in the Y3 analysis. This paper is one of the two dedicated to the analysis of the BAO signal in DES Y6. In its companion paper, we present the angular diameter distance constraints obtained through the fitting to the BAO scale.

%% file: acknowledgements.tex
\textit{Author’s contributions}: We would like to acknowledge everyone who made this work possible.
JMF developed the code to optimize the sample selection, performed the redshift calibration, developed the shift and stretch fitting code and ran the tests on $w(\theta)$.
MRM conducted the observational systematics analysis, creating the angular mask, obtaining systematic weights with the ISD method and running the corresponding validation tests.
SA coordinated and overviewed the BAO team and project, as well as the development of this manuscript. 
APo coordinated the BAO project and supervised the sample optimization. 
KCC performed the tests on $\xi_{\rm p}$. 
HC performed the tests on $C_\ell$.
NW contributed to the definition of the mask, SP maps, weights and stellar contamination estimates.
ISN worked on the curation and validation of the sample, provided information on the raw and value-added data and survey maps and participated in general discussions at the early and mid stages of the project.
ESa contributed to the supervision of JMF and the photo-$z$ validation and also provided critical feedback that helped shape the analysis. 
LTSC participated in the estimation of photo-$z$ using the DNF algorithm, particularly in the selection of the training sample and the validation of the results.
JDV adapted the DNF photometric redshift code to the requirements of the BAO sample.
RC did the clustering redshift measurements.
ACR created the original Y6 angular mask and contributed as an internal reviewer.
JEP contributed to the development of the observational systematics pipeline and as an internal reviewer.
GG provided the SOMPZ redshift distributions. SL created simulations for the BAO sample to validate the observational systematic pipeline. 
MA, KB, ADW, RAG, WGH, APi, ESR, ESh and BY contributed to the creation of the Y6 Gold catalog. AJR provided the original version of the Fisher forecast code.
The remaining authors have made contributions to this paper that include, but are not limited to, the construction of DECam and other aspects of collecting the data; data processing and calibration; developing broadly used methods, codes and simulations; running the pipelines and validation tests; and promoting the science analysis.

APo acknowledges support from the European Union’s Horizon Europe program under the Marie Skłodowska-Curie grant agreement 101068581. KCC is supported by the National Science Foundation of China under the grant number 12273121 and the science research grants from the China Manned Space Project. HC is supported by FAPESP and CNPq.


Funding for the DES Projects has been provided by the U.S. Department of Energy, the U.S. National Science Foundation, the Ministry of Science and Education of Spain, 
the Science and Technology Facilities Council of the United Kingdom, the Higher Education Funding Council for England, the National Center for Supercomputing 
Applications at the University of Illinois at Urbana-Champaign, the Kavli Institute of Cosmological Physics at the University of Chicago, 
the Center for Cosmology and Astro-Particle Physics at the Ohio State University,
the Mitchell Institute for Fundamental Physics and Astronomy at Texas A\&M University, Financiadora de Estudos e Projetos, 
Funda{\c c}{\~a}o Carlos Chagas Filho de Amparo {\`a} Pesquisa do Estado do Rio de Janeiro, Conselho Nacional de Desenvolvimento Cient{\'i}fico e Tecnol{\'o}gico and 
the Minist{\'e}rio da Ci{\^e}ncia, Tecnologia e Inova{\c c}{\~a}o, the Deutsche Forschungsgemeinschaft and the Collaborating Institutions in the Dark Energy Survey. 

The Collaborating Institutions are Argonne National Laboratory, the University of California at Santa Cruz, the University of Cambridge, Centro de Investigaciones Energ{\'e}ticas, 
Medioambientales y Tecnol{\'o}gicas-Madrid, the University of Chicago, University College London, the DES-Brazil Consortium, the University of Edinburgh, 
the Eidgen{\"o}ssische Technische Hochschule (ETH) Z{\"u}rich, 
Fermi National Accelerator Laboratory, the University of Illinois at Urbana-Champaign, the Institut de Ci{\`e}ncies de l'Espai (IEEC/CSIC), 
the Institut de F{\'i}sica d'Altes Energies, Lawrence Berkeley National Laboratory, the Ludwig-Maximilians Universit{\"a}t M{\"u}nchen and the associated Excellence Cluster Universe, 
the University of Michigan, NSF's NOIRLab, the University of Nottingham, The Ohio State University, the University of Pennsylvania, the University of Portsmouth, 
SLAC National Accelerator Laboratory, Stanford University, the University of Sussex, Texas A\&M University and the OzDES Membership Consortium.

Based in part on observations at Cerro Tololo Inter-American Observatory at NSF's NOIRLab (NOIRLab Prop. ID 2012B-0001; PI: J. Frieman), which is managed by the Association of Universities for Research in Astronomy (AURA) under a cooperative agreement with the National Science Foundation.

The DES data management system is supported by the National Science Foundation under Grant Numbers AST-1138766 and AST-1536171.
The DES participants from Spanish institutions are partially supported by MICINN under grants PID2021-123012, PID2021-128989 PID2022-141079, SEV-2016-0588, CEX2020-001058-M and CEX2020-001007-S, some of which include ERDF funds from the European Union. IFAE is partially funded by the CERCA program of the Generalitat de Catalunya. Research leading to these results has received funding from the European Research
Council under the European Union's Seventh Framework Program (FP7/2007-2013) including ERC grant agreements 240672, 291329 and 306478.
We acknowledge support from the Brazilian Instituto Nacional de Ciencia
e Tecnologia (INCT) do e-Universo (CNPq grant 465376/2014-2).


This manuscript has been authored by Fermi Research Alliance, LLC under Contract No. DE-AC02-07CH11359 with the U.S. Department of Energy, Office of Science, Office of High Energy Physics.

This paper uses data from the VIMOS Public Extragalactic Redshift Survey (VIPERS). VIPERS has been performed using the ESO Very Large Telescope, under the ``Large Programme'' 182.A-0886. The participating institutions and funding agencies are listed at \url{http://vipers.inaf.it}.

This paper is part of a series of two papers that measure the baryon acoustic oscillations signal using the Dark Energy Survey final dataset, with \cite{Y6_BAO_measurement} being its companion paper.


